\documentclass[twocolumn,prb,showpacs,floatfix,amsfonts,nofootinbib]{revtex4-1}
\pdfoutput=1
\usepackage[latin9]{inputenc}
\usepackage{amsmath}
\usepackage{amssymb}
\usepackage{graphicx}
\usepackage{esint}

\makeatletter
\usepackage[pdftex]{hyperref}% Include figure files

\makeatother

\begin{document}

\title{Scattering of surface plasmon-polaritons in a graphene multilayer
photonic crystal with inhomogeneous doping}

\author{Yu. V. Bludov, N. M. R. Peres, G. Smirnov, M. I. Vasilevskiy }

\affiliation{Department of Physics and Center of Physics, University of Minho,
P-4710-057, Braga, Portugal}

%\date{\today}
\begin{abstract}
The propagation of a surface plasmon-polariton  along a stack of doped graphene sheets is considered.
This auxiliary problem is used to discuss: (i) the scattering of such a mode at an interface between the stack and
the vacuum; (ii) the scattering at an interface where there is a sudden change of the electronic doping. The formalism
is then extended to the {\it barrier problem}. In this system rich physics is found for the plasmonic mode, showing:
total reflection, total transmission, Fabry-P\'erot oscillations, and coupling to photonic modes.
\end{abstract}


\maketitle

\section{Introduction}

Plasmonics deals with the excitation, manipulation, and
utilization of surface plasmon-polaritons (SPPs), where the latter
are hybridized excitations of radiation with the collective charge oscillations
of an electron gas \cite{Maier,ACSgp,MRS:8669483}. In traditional noble-metal plasmonics the electron gas is provided by the
free electrons in the metal. Furthermore, SPPs are excited at the interface
between a metal and a dielectric and propagate along the interface with exponential localization in the
direction perpendicular to that of their motion.

One central idea in plasmonics is to explore the sub-wavelength
confinement of light to build plasmonic waveguides that would propagate, at the same time, an electric signal
and a highly confined electromagnetic wave \cite{Stockman}.
A plasmonic circuitry would involve lenses, mirrors, beam splitters,
and the like. Therefore, the study of scattering of plasmons by such structures arises.

Clearly, the problem of scattering of plasmons is a scientific and technological one.
The deep understanding of the scattering of plasmons is instrumental for building new technologies.
In traditional noble-metal plasmonics, the range of wavelengths where SPPs
show sub-wavelength confinement is restricted to the interval spanning the near infrared (near-IR) to the ultraviolet.
In the mid-infrared (mid-IR) to the terahertz (THz) spectral range SPPs in structures with noble metals are essentially free radiation,
therefore lacking the key advantage of sub-wavelength confinement. This makes them unsuitable for
plasmonics circuitry and sensing \cite{MRS:8669489}.

From what has been said above it follows that new plasmonic materials able of showing sub-wavelength
confinement and spanning the frequency interval ranging from the THz to the mid-IR are necessary. This is particularly relevant
as important biomolecules exhibit unique spectral signatures in this frequency range. Thus, the sensing
capability arising with noble-metal plasmonics in the near-IR to the ultraviolet
could be extended to a region that the traditional systems cannot cover. Such possibility would increase the
application of plasmonics for sensing and security applications, such as detection of pollutants, diagnosis
of diseases, food control quality, and detection of plastic explosives.

It is in the above context that graphene
emerges
as a promising plasmonic material \cite{AbajoACSP,c:Bludov2013,JPCM26,nanoImICFO,nanoImBasov,ACSgp,Andre}.
SPPs in graphene
exist in the THz to mid-IR range and show a high degree of sub-wavelength localization, therefore circumventing the
mentioned limitations of noble-metal plasmonics. Indeed, it can be shown that the degree of localization of plasmons in graphene
is given by \cite{nlgp}
\begin{equation}
 \zeta_G\propto \alpha \hbar c \frac{E_F}{(\hbar\omega)^2}\,,
\end{equation}
where $\alpha$ is the fine structure constant of atomic physics, $c$ is the speed of light, $E_F$ is the Fermi energy of graphene,
and $\omega$ is the frequency of the surface plasmon-polariton. Taking, as an example, a frequency of 150 THz
(equivalent to the wavelength of $\lambda_0=$2 $\mu$m for the radiation in vacuum, which corresponds to the edge of the mid-IR region),
and considering a typical Fermi energy of 0.5 eV
(a value easily attainable by the electrostatic gating) we obtain for $\zeta_G\sim0.002$ $\mu$m, that is
\begin{equation}
 \frac{\lambda_0}{\zeta_G}\sim 10^3\,,
\end{equation}
which is a rather high degree of localization. This value yields highly intense and localized
electromagnetic fields. The above estimation highlights the potential of graphene plasmonics in the THz to mid-IR
spectral range.

Being a two-dimensional membrane, graphene is amenable for stacking. The idea is to build a photonic crystal
composed by several stacked sheets of graphene separated by  dielectric layers;
this structure has been investigated both theoretically
\cite{c:Berman2010,c:Berman2012,c:Arefinia2013,c:Qin2014,PhysRevB.92.195425,c:Hajian2013,c:Bludov2013,c:El-Naggar2015,c:Al-sheqefi2015,c:Madani2013,c:Cheng2015,c:Kaipa2012}
and experimentally\cite{c:Xu-experim,c:Sreekanth2012}.
In such  structures charge carriers in different graphene layers are able to interact
by means of electromagnetic waves, which can be either propagating or evanescent inside the dielectric.
The latter case refers to the area of plasmonics, where interaction between the SPPs supported by each of the
graphene layers results in the formation of polaritonics
bands\cite{c:Hajian2013,c:Bludov2013,c:Fan_plasmon2013,c:Wang_plasmon2012}.
This fact allows for the existence of a number of interesting phenomena such as Bloch\cite{c:Fan_Bloch2014} and Rabi\cite{c:Wang_Rabi2015}
oscillations of SPPs as well as the formation of nonlinear self-localized wavepackets
---lattice solitons\cite{c:Bludov_solitons2015,c:Wang_solitons2015}.
Moreover, it was predicted that a plasmonic biosensor based on a graphene multilayer system
shows a much higher sensitivity than its counterpart operating using a gold film\cite{c:Sreekanth_biosensor2013}.
The propagation of bulk waves in a graphene stack is characterized by several phenomena typical for
periodic structures, like the presence of the omnidirectional low-frequency gap\cite{c:El-Naggar2015,c:Al-sheqefi2015,c:Madani2013}
in the spectrum (which is not present in the photonic crystals without graphene), extraordinary
absorption decrease\cite{c:Cheng2015}, and light pulse delay \cite{c:Liu2014}.
%The bulk wave spectrum of the graphene stack is characterized by the presence of the onnidirectional low-frequency gap\cite{c:El-Naggar2015,c:Al-sheqefi2015,c:Madani2013} as well as by extraordinary absorption decrease\cite{c:Cheng2015}.  in the  in graphene stacks is influenced by the   The
In practice, these graphene multilayers can be used as terahertz modulators\cite{c:Sensale-Rodriguez2012},
broadband polarizers\cite{c:Yan2012}, tunable Bragg reflectors \cite{c:Bludov2013},
and polarization splitters \cite{c:Al-sheqefi2015}. Also it is interesting that the
graphene stacks exhibit the properties of hyperbolic
metamaterials \cite{c:Iorsh2013,c:Zhukovsky2014,c:Cheng_metamaterial2015}.
Finally, the study of propagation of radiation in a disordered graphene stack when light impinges perpendicularly
to the graphene surface has recently been considered \cite{PhysRevB.92.195425}, showing that graphene can control Anderson localization
of radiation.

As mentioned above, eventually it will be necessary to build some kind of plasmonic
circuitry where the problem of scattering of SPPs arises.
In graphene, it is possible to control the percentages of reflection and transmittance of
a surface plasmon-polariton by controlling the local value of the electronic density.
Consider the  simplest case of a graphene sheet on a split gate. Each of the two parts of
the gate is subjected to different gate potentials, and this creates two zones in
the material presenting two different electronic concentrations. Assuming now that a surface
plasmon-polariton is impinging on the border line defined by the split gate, the amount of power
reflected and transmitted will be controlled by the difference in the local electronic densities.
The problem just described has already been discussed in the literature \cite{c:Vakil2011,KhavasiScatt}.
Another interesting question is the coupling of SPPs to photonic modes.
The idea can work in two ways: either a photonic mode will excite a surface plasmon-polariton in
graphene or a surface plasmon-polariton propagating on graphene will radiate to free space as a
photonic mode. Both cases find relevant technological applications.

As we will show in the bulk of the article, it is easier to achieve
the interaction between SPP and photonic modes in the graphene stack, than in single graphene layer. Due to the periodicity of the system, this interaction can be direct (i.e. without using prisms).
Also the scattering of SPPs in a stack of graphene sheets has rich physics, including total reflection, total transmission, and Fabry-P\'erot oscillations.

%Experimentally it was demonstrated that in graphene multilayer \cite{c:Xu-experim} experiment \cite{c:Sreekanth2012}

%it was proposed to use the graphene multilayer as terahertz modulators\cite{c:Sensale-Rodriguez2012}, broadband polarizer\cite{c:Yan2012}, and tunable Bragg reflectors \cite{c:Bludov2013}photonic crystals \cite{c:Madani2013}.

 %polarization splitter \cite{c:Al-sheqefi2015}.

The paper is organized as follows. In Sec.\ref{sec:eigen} we consider
the eigenvalues and the eigenfunctions of a graphene multilayer photonic crystal (PC), which
will serve as a basis for the following sections. Sec.\ref{sec:single-interface}
is devoted to two problems: (i) scattering of an incident polaritonic
mode at the interface between the graphene multilayer PC and a homogeneous
dielectric;  (ii) scattering  of an incident polaritonic
mode at the interface between two PCs, characterized by
different Fermi energies of the graphene sheets.
In Sec.\ref{sec:double-interface}
we consider the scattering  of a polaritonic mode on a double interface
between PCs.

\section{Eigenmodes of the graphene multilayer photonic crystal}

\label{sec:eigen}
\begin{figure*}
\includegraphics[width=16cm]{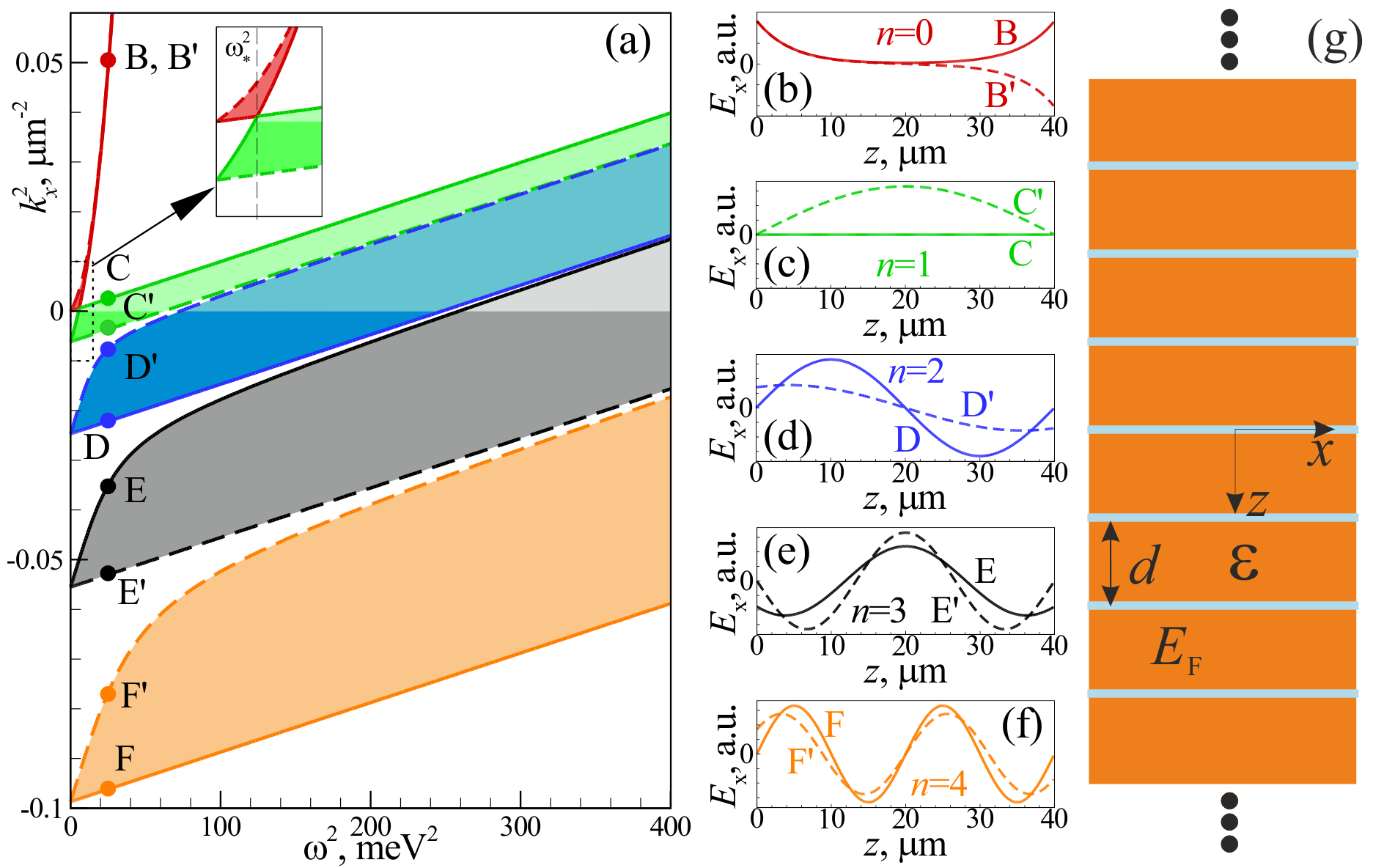}

\caption{(a) Dispersion curves (squared $x-$component of wavevector $k_{x}^{2}$
\emph{versus} squared frequency $\omega^2$) of graphene multilayer PC: colored
domains correspond to the allowed bands, which boundaries are determined
by Bloch wavevector at center of Brillouin zone $q=0$ (bold solid
lines, like B--F), or at its edge $q=\pi/d$ (bold dashed lines, like
B'--F'); (b)--(f) Spatial profiles of the multilayer graphene PC's
eigenfunctions at $\omega=5\,\mathrm{meV}$. The parameters, which
correspond to each of the eigenfunctions in panels (b)--(f), are depicted
by respective points B--F, B'--F' in panel (a). In all panels other
parameters are: $d=40\,\mu$m, $E_{F}=0.157\,$eV (which
correspond to the gate voltage 25 V, applied to graphene on top of
the 300 nm thickness $\mathrm{SiO}{}_{2}$ substrate), $\varepsilon=3.9$,
$\Gamma=0$; (g) Schematic view of the graphene-based PC.}\label{fig:sl-p}
\end{figure*}

In order to calculate the reflection of SPPs from the interface between two graphene multilayer PCs,
it is necessary to know the spectrum of the electromagnetic waves
in PC. In the present section we consider an auxiliary problem of eigenmodes in
a multilayer graphene stack composed of an infinite number of single
graphene layers with equal Fermi energy $E_{F}$ {[}see Fig.\ref{fig:sl-p}(g){]}.
We suppose that graphene layers are arranged at equal distances $d$
from each other at planes $z=md$, $m\in\left(-\infty,\infty\right)$
and are embedded into a uniform dielectric medium with a dielectric
constant $\varepsilon$. If the electromagnetic field is uniform along
the direction $y$ ($\partial/\partial y\equiv0$), then it can be
decomposed into two separate waves of different polarizations.
In the following we restrict our consideration to $p$-polarized
waves, whose magnetic field is perpendicular to the plane of
incidence ($xz$). Such a wave possesses the electromagnetic field components
$\vec{E}=\left\{ E_{x},0,E_{z}\right\} ,$ $\vec{H}=\left\{ 0,H_{y},0\right\} $,
and is described by the Maxwell equations
\begin{eqnarray}
\frac{\partial E_{x}}{\partial z}-\frac{\partial E_{z}}{\partial x}=i\kappa H_{y},\label{eq:max1-p}\\
\frac{\partial H_{y}}{\partial z}=i\kappa\varepsilon E_{x},\qquad\frac{\partial H_{y}}{\partial x}=-i\kappa\varepsilon E_{z}.\label{eq:max2-p}
\end{eqnarray}
Here we assumed the temporal dependence of the electromagnetic field
$\vec{E},\vec{H}$ in the form of $\exp(-i\omega t)$, where $\omega$
is the cyclic frequency, $\kappa=\omega/c$ and $c$ is the speed
of light in vacuum. The electromagnetic properties
of the graphene layer are determined by its dynamical conductivity
$\sigma_{g}\left(\omega\right)$, whose form can be found, e.g., in
Ref.\cite{c:primer}. In order to find the dispersion relation of
the graphene multilayer PC, the electromagnetic fields should be considered
separately in each layer between the adjacent graphene sheets at planes
$z=md$ and $z=(m+1)d$. The solutions of the Maxwell equations
(\ref{eq:max1-p})--(\ref{eq:max2-p}) at the spatial domain $md\leq z\leq(m+1)d$
can be represented as
\begin{eqnarray}
H_{y}(x,z)=\left\{ H_{m,+}\exp\left[ik_{z}\left(z-md\right)\right]+\right.\label{eq:hy-sl}\\
\left.H_{m,-}\exp\left[-ik_{z}\left(z-md\right)\right]\right\} \exp(ik_{x}x),\nonumber \\
E_{x}(x,z)=\frac{k_{z}}{\kappa\varepsilon}\left\{ H_{m,+}\exp\left[ik_{z}\left(z-md\right)\right]-\right.\label{eq:ex-sl}\\
\left.H_{m,-}\exp\left[-ik_{z}\left(z-md\right)\right]\right\} \exp(ik_{x}x)\nonumber \\
E_{z}(x,z)=-\frac{k_{x}}{\kappa\varepsilon}\left\{ H_{m,+}\exp\left[ik_{z}\left(z-md\right)\right]+\right.\label{eq:ez-sl}\\
\left.H_{m,-}\exp\left[-ik_{z}\left(z-md\right)\right]\right\} \exp(ik_{x}x).\nonumber
\end{eqnarray}
Here $k_{x}$ is the in-plane component of the wavevector
(parallel to the graphene sheets), $k_{z}=\left(\kappa^{2}\varepsilon-k_{x}^{2}\right)^{1/2}$,
$H_{m,\pm}$ are the amplitudes of the forward (sign ''+'') or backward
(sign \textquotedblright --\textquotedblright ) propagating waves.
By matching boundary conditions at $z=md$ [continuity of the tangential
component of the electric field across the graphene $E_{x}(x,md+0)=E_{x}(x,md-0)$,
and discontinuity of the tangential component of the magnetic field,
caused by surface currents in graphene, $H_{y}(x,md+0)-H_{y}(x,md-0)=-(4\pi/c)j_{x}=-(4\pi/c)\sigma_{g}E_{x}(x,md)$],
one can find that amplitudes $H_{m,\pm}$ can be related to $H_{m-1,\pm}$
as

\begin{eqnarray}
\left(\begin{array}{c}
H_{m,+}\\
H_{m,-}
\end{array}\right)=\hat{M}\left(\begin{array}{c}
H_{m-1,+}\\
H_{m-1,-}
\end{array}\right),\label{eq:h-mp}
\end{eqnarray}
where the matrix $\hat{M}$ reads as
\begin{eqnarray*}
\hat{M}=\left(\begin{array}{cc}
\exp\left(ik_{z}d\right)\left[1-i\Lambda k_{z}\right] & i\Lambda k_{z}\exp\left(-ik_{z}d\right)\\
-i\Lambda k_{z}\exp\left(ik_{z}d\right) & \exp\left(-ik_{z}d\right)\left[1+i\Lambda k_{z}\right]
\end{array}\right)
\end{eqnarray*}
with $\Lambda=2\pi\sigma_{g}/(i\omega\varepsilon)$. Since the considered
structure is periodic, it is possible to use the Bloch theorem, which
determines the proportionality between field amplitudes in the adjacent
periods through the Bloch wavevector $q$:
\begin{eqnarray}
H_{m-1,\pm} & = & \exp\left(-iqd\right)H_{m,\pm},\label{eq:Bloch-rel}
\end{eqnarray}
After substitution of this relation into Eqs.(\ref{eq:h-mp}), the
solvability condition of the resulting linear equations requires (here
$\hat{I}$ is the unit matrix)
\begin{eqnarray}
\mathrm{Det}\left|\hat{M}-\exp\left(iqd\right)\hat{I}\right|=0,\label{eq:disp-p-mat}
\end{eqnarray}
which results into the dispersion relation for the $p$-wave in graphene
multilayer PC
\begin{equation}
\cos\left(qd\right)-\cos\left(k_{z}d\right)-\Lambda k_{z}\sin\left(k_{z}d\right)=0.\label{eq:disp-p}
\end{equation}
Similar expressions for the dispersion relation were obtained in Ref.\cite{c:Hajian2013,c:Bludov2013}. The dispersion relation Eq.(\ref{eq:disp-p}) for fixed $\omega$ and $q$ possesses an infinite number of solutions for $k_x$. Further in the paper we will prescribe an index $n\ge 0$ as a superscript to all parameters distinguishing the respective eigenmode. We also note, that the solvability conditions for
Eq.(\ref{eq:disp-p-mat}) [as well as the dispersion relation (\ref{eq:disp-p})]
imply a simple relation between forward- and backward propagating
waves
\begin{eqnarray}
\frac{H_{m,+}^{(n)}}{H_{m,-}^{(n)}}=-\frac{\exp\left(-ik_{z}^{(n)}d\right)\Lambda k_{z}^{(n)}}{\exp\left(ik_{z}^{(n)}d\right)\left[1-\Lambda k_{z}^{(n)}\right]-\exp\left(iqd\right)}=\nonumber\\
=\frac{\exp\left(iqd\right)-\exp\left(-ik_{z}^{(n)}d\right)}{\exp\left(iqd\right)-\exp\left(ik_{z}^{(n)}d\right)},\nonumber
\end{eqnarray}
which allows to represent these amplitudes as
\begin{eqnarray}
H_{m,+}^{(n)}={\cal H}^{(n)}\left(q,\omega\right)\frac{\exp\left(iqd\right)-\exp\left(-ik_{z}^{(n)}d\right)}{2\sqrt{A^{(n)}}}\exp\left(iqmd\right),\nonumber\\ \label{eq:Hp}\\
H_{m,-}^{(n)}={\cal H}^{(n)}\left(q,\omega\right)\frac{\exp\left(iqd\right)-\exp\left(ik_{z}^{(n)}d\right)}{2\sqrt{A^{(n)}}}\exp\left(iqmd\right),\nonumber\\
\label{eq:Hm}
\end{eqnarray}
where ${\cal H}^{(n)}\left(q,\omega\right)$ is the magnetic field amplitude and $A^{(n)}$ is a normalization factor. Notice
that amplitudes being represented in this form also satisfy Bloch
condition (\ref{eq:Bloch-rel}). Substituting Eqs.(\ref{eq:Hp}) and
(\ref{eq:Hm}) into (\ref{eq:hy-sl}), we obtain the expression for
the component of the electromagnetic field at spatial domain $md\leq z\leq(m+1)d$ in the
form
\begin{eqnarray}
H_{y,\pm}^{(n)}(x,z||q,\omega)={\cal H}_{\pm}^{(n)}\left(q,\omega\right)\times\nonumber\\
\psi^{(n)}\left(z||q,\omega\right)\exp(\pm ik_{x}^{(n)}x),\label{eq:eigen-Hyn}\\
E_{x,\pm}^{(n)}(x,z||q,\omega)=\frac{{\cal H}_{\pm}^{(n)}\left(q,\omega\right)}{i\kappa\varepsilon}\times\nonumber\\
\frac{\partial\psi^{(n)}\left(z||q,\omega\right)}{\partial z}\exp(\pm ik_{x}^{(n)}x),\label{eq:eigen-Exn}\\
E_{z,\pm}^{(n)}(x,z||q,\omega)=\mp{\cal H}_{\pm}^{(n)}\left(q,\omega\right)\frac{k_{x}^{(n)}}{\kappa\varepsilon}\times\nonumber\\
\psi^{(n)}\left(z||q,\omega\right)\exp(\pm ik_{x}^{(n)}x),\label{eq:eigen-Ezn}
\end{eqnarray}
where
\begin{eqnarray}
\psi^{(n)}\left(z||q,\omega\right)=\left\{ \exp\left[iqd\right]\cos\left[k_{z}^{(n)}\left(z-md\right)\right]\right.\label{eq:psi}\\
\left.-\cos\left[k_{z}^{(n)}\left(md+d-z\right)\right]\right\}\frac{\exp\left[iqmd\right]}{\sqrt{A^{(n)}}},\nonumber
\end{eqnarray}
 is a dimensionless spatial profile function. Here the normalization factor
\begin{eqnarray}
A^{(n)}=1-\cos\left(qd\right)\cos\left(k_{z}^{(n)}d\right)+\nonumber\\
\frac{\cos\left(k_{z}^{(n)}d\right)-\cos\left(qd\right)}{k_{z}^{(n)}d}\sin\left(k_{z}^{(n)}d\right) \nonumber
\end{eqnarray}
is chosen to satisfy the condition
\[
\frac{1}{d}\int_{md}^{md+d}\left|\psi^{(n)}\left(z||q,\omega\right)\right|^{2}dz=1.
\]
Also we take into account that all eigenmodes of the PC can be either
forward- or backward propagating: this fact is stressed in Eqs.(\ref{eq:eigen-Hyn})--(\ref{eq:eigen-Ezn})
by adding the signs ``+'' or ``-'' before the $x$-component of
wavevector $k_{x}^{(n)}$ as well as by the subscript in the amplitude
${\cal H}_{\pm}^{(n)}$.

Before considering the dispersion properties in detail, it should
be noticed that further in the paper we neglect collisional losses and take the relaxation rate in graphene $\Gamma=0$, i.e., the graphene
conductivity is supposed to be purely imaginary. This
approximation simplifies the analysis without changing quantitatively
the results. The spatial periodicity of the multilayer graphene PC gives
rise to the band-gap structure {[}see Fig.\ref{fig:sl-p}(a){]}: the
spectrum is composed of an infinite number of bands {[}$n\ge0$,
colored domains in Fig.\ref{fig:sl-p}(a){]}, whose boundaries are
determined by the Bloch wavevector at the center $q=0$, or at edge
$q=\pi/d$ of Brillouin zone {[}bold solid and bold dashed lines in
Fig.\ref{fig:sl-p}(a), respectively{]}.
One of the bands {[}$n=0$, depicted by red
color in Fig.\ref{fig:sl-p}(a){]} is polaritonic, where
electromagnetic waves are evanescent in $z-$direction (with purely
imaginary $k_{z}^{(0)}$), and propagating in $x-$direction (with
purely real $k_{x}^{(0)}$). The other bands ($n\ge1$) are photonic
ones, four of which with $n=1,...,4$ are depicted in Fig.\ref{fig:sl-p}(a)
by green, blue, black and orange colors, respectively. All photonic
bands are characterized by the propagating nature of the electromagnetic
waves in $z-$direction (with purely real $k_{z}^{(n)}$), while in
$x$-direction they can be either propagating
{[}$\left(k_{x}^{(n)}\right)^{2}>0$, shaded by lighter colors in Fig.\ref{fig:sl-p}(a){]}, or evanescent {[}$\left(k_{x}^{(n)}\right)^{2}<0$, shaded by darker colors in Fig.\ref{fig:sl-p}(a){]}.
The latter are physically meaningful only in confined photonic crystals because they diverge in either plus or minus infinity.
Notice that, as follows from the inset in Fig.\ref{fig:sl-p}(a), the polaritonic and the first photonic
band touch each other at a cutoff frequency $\omega_{*}\approx\left(4\alpha E_{F}c/\hbar\varepsilon d\right)^{1/2}$ in the center of the Brillouin zone, $q=0$ (see Appendix A for details). Below this frequency, for $\omega<\omega_{*}$,
the edge of the polaritonic band at $q=0$ coincides with the light
line $k_{x}^{2}=\omega^{2}\varepsilon/c^{2}$, while above it, for $\omega>\omega_{*}$, the boundary of the polaritonic band
$q=0$ detachs from the light line {[}which coincides with the edge of the first photonic band, solid
green line in Fig.\ref{fig:sl-p}(a){]}. At high frequencies, when the 
localization of the polaritonic modes near each graphene layer is strong, and
the SPPs, sustained by neighboring graphene layers, are
almost noninteracting. This fact gives rise to the situation when
polaritonic dispersion curves for different Bloch wavevectors $q$
merge together {[}see e.g. points B and B' in Fig.\ref{fig:sl-p}(a){]},
although their spatial profiles remain different, as it is evident
from Fig.\ref{fig:sl-p}(b).

Some of the edges of the photonic bands are characterized by an interesting
property: at $q=\pi/d$ the edges of the photonic bands with odd numbers,
$n=2l-1$ ($1\le l<\infty$) possesses $z$-components of the wavevector
\begin{equation}
k_{z}^{(2l-1)}=\left\{ \frac{\omega}{c^{2}}^{2}\varepsilon-\left(k_{x}^{(2l-1)}\right)^{2}\right\} ^{1/2}=\frac{2l-1}{d}\pi.\label{eq:w-g-pi}
\end{equation}
Examples of such modes are C' and E' in Figs.\ref{fig:sl-p}(c) and
\ref{fig:sl-p}(e). Corresponding expressions for the eigenfunctions
(\ref{eq:psi}) can be represented as (details can be found in Appendix
A)
\begin{eqnarray}
\psi^{(2l-1)}\left(z||\frac{\pi}{d},\omega\right)=\sqrt{2}\cos\left[\frac{2l-1}{d}\pi z\right].\label{eq:shape-w-pi}
\end{eqnarray}
When $q=0$, the edges of the bands with even numbers $n=2l$ ($1\le l<\infty$)
possess the same property, i.e.
\begin{eqnarray}
k_{z}^{(2l)}=\left\{ \frac{\omega}{c^{2}}^{2}\varepsilon-\left(k_{x}^{(2l)}\right)^{2}\right\} ^{1/2}=\frac{2l}{d}\pi,\label{eq:w-g-0}\\
\psi^{(2l)}\left(z||0,\omega\right)=\sqrt{2}\cos\left[\frac{2l}{d}\pi z\right].\label{eq:shape-w-0}
\end{eqnarray}
Examples of such modes are D and F in Figs.\ref{fig:sl-p}(d) and
\ref{fig:sl-p}(f). It is interesting that expressions for solutions
(\ref{eq:w-g-pi}), (\ref{eq:w-g-0}) do not contain the graphene
conductivity $\sigma_{g}$ (in other words, their spectrum does not
depend upon the graphene's Fermi energy $E_{F}$):
$z$-components $k_{z}^{(2l)}$ and $k_{z}^{(2l-1)}$,
specified by Eqs.(\ref{eq:w-g-pi}) and (\ref{eq:w-g-0}), cancel
conductivity-dependent term in the dispersion relation (\ref{eq:disp-p}).
This happens because the derivatives of their spatial profile functions vanish, $\partial\psi^{(n)}/\partial z$=0
at $z=md$ {[}see (\ref{eq:shape-w-pi}), (\ref{eq:shape-w-0}){]}.
According to (\ref{eq:eigen-Exn}), it implies zero
tangential components of the electric field $E_{x,\pm}^{(n)}(x,md||q)=0$
at graphene layers, clearly seen from the spatial profiles of the modes C',
D, E', F in Figs.\ref{fig:sl-p}(c)--\ref{fig:sl-p}(f). The light
line $k_{x}=\omega\sqrt{\varepsilon}/c$ mentioned above is also an
eigenmode of the multilayer graphene PC with $q=0$ {[}see mode C in
Fig.\ref{fig:sl-p}(c){]}. It is characterized by absence of the wavevector's
$z$-component $k_{z}^{(n)}=0$ and the spatial profile function $\psi^{(n)}\left(z||0,\omega\right)=1$
(further in the paper this mode will be referred to as the \textit{light-line mode}).
The band index for this mode is $n=0$ in the frequency domain $\omega<\omega_{*}$
(where this mode is the edge of polaritonic band) and $n=1$ above the
critical frequency $\omega>\omega_{*}$(where this mode is the edge
of the first photonic band). As a matter of fact, this mode is a bulk
electromagnetic wave propagating in the $x$-direction and with zero
longitudinal component of the electric field $E_{x}^{(n)}(x,z||q,\omega)\equiv0$,
i.e. a purely transverse wave.

\section{Scatttering of polaritonic mode from a single interface between
two graphene multilayer PCs }

\label{sec:single-interface}
\begin{figure}
\includegraphics[width=8.5cm]{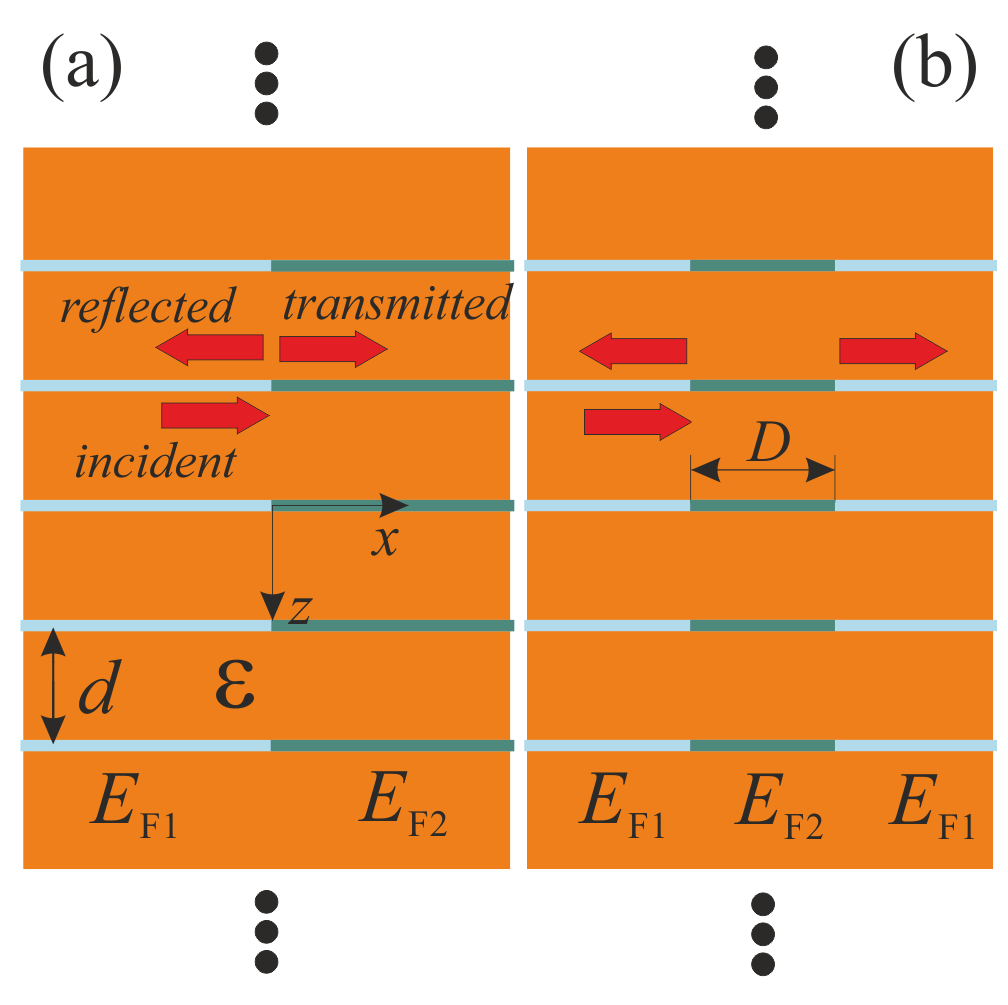}

\caption{A single {[}panel (a){]} or double {[}panel (b){]} interface between
two graphene multilayer PCs with different Fermi energies of graphene
layers.\label{fig:geometry_step_barrier}}
\end{figure}

Let us consider now an interface between
two graphene multilayer PCs (described in Sec.\ref{sec:eigen}). We
suppose that both PCs possess the same period $d$, are embedded
in the same homogeneous dielectric medium with the dielectric permittivity
$\varepsilon$, and graphene layers are arranged in the same planes
$z=md$ along axis $z$. The only difference between these graphene
multilayer PCs is the Fermi energy, which is equal to $E_{F1}$ in
the first PC (occupying the half-space $x<0$) and to $E_{F2}$ in
the second one (occupying the half-space $x>0$). Further in the text the left and right PCs will be referred to as PC1 and PC2, respectively. Such an interface
is schematically depicted in Fig.\ref{fig:geometry_step_barrier}.
We consider the situation, where the polaritonic mode propagates in
the positive direction of the $x$-axis (thus coming from $x=-\infty$)
and impinges on the aforementioned interface. The main objective of
the present section is to describe the scattering of the polaritonic
mode on this interface, that is, we want to find the transmission
and reflection coefficients of the polaritonic mode as well as to
determine which part of its energy is transferred to each of the photonic modes.

In order to do this, we expand the electromagnetic field in series
with respect to the graphene multilayer PC eigenmodes (\ref{eq:eigen-Hyn})--(\ref{eq:eigen-Ezn}).
So, the magnetic field and the $z$-component of the electric field in
PC1 ($x<0$) can be written as
\begin{eqnarray}
H_{1,y}(x,z)={\cal H}_{1,+}^{(0)}\psi_{1}^{(0)}\left(z\right)\exp(ik_{1,x}^{(0)}x)+\nonumber\\
\sum_{n=0}^{\infty}{\cal H}_{1,-}^{(n)}\psi_{1}^{(n)}\left(z\right)\exp(-ik_{1,x}^{(n)}x),\label{eq:Hy-1} \\
E_{1,z}(x,z)=-{\cal H}_{1,+}^{(0)}\frac{k_{1,x}^{(0)}}{\kappa\varepsilon}\psi_{1}^{(0)}\left(z\right)\exp(ik_{1,x}^{(0)}x)+\nonumber\\
\sum_{n=0}^{\infty}{\cal H}_{1,-}^{(n)}\frac{k_{1,x}^{(n)}}{\kappa\varepsilon}\psi_{1}^{(n)}\left(z\right)\exp(-ik_{1,x}^{(n)}x).\label{eq:Ez-1}
\end{eqnarray}
In the same manner, the fields in PC2 can be expressed as
\begin{eqnarray}
H_{2,y}(x,z)=\sum_{l=0}^{\infty}{\cal H}_{2,+}^{(l)}\psi_{2}^{(l)}\left(z\right)\exp(ik_{2,x}^{(l)}x),\label{eq:Hy-2}\\
E_{2,z}(x,z)=-\sum_{l=0}^{\infty}{\cal H}_{2,+}^{(l)}\frac{k_{2,x}^{(l)}}{\kappa\varepsilon}\psi_{2}^{(l)}\left(z\right)\exp(ik_{2,x}^{(l)}x)\; .
\label{eq:Ez-2}
\end{eqnarray}
In Eqs.(\ref{eq:Hy-1})--(\ref{eq:Ez-2}) we prescribe the  PC index $j=1,2$ to the spatial profile function $\psi_{j}^{(n)}$,
the $x$-component of wavevector $k_{j,x}^{(n)}$, and the amplitude
${\cal H}_{j,\pm}^{(n)}$. Also we use band indices
$n$ and $l$, referring to PC1 and PC2, respectively, and drop the arguments $q$ and $\omega$ (which are equal for PC1 and PC2) in functions and amplitudes for brevity.
Notice that Eqs.(\ref{eq:Hy-1}) and (\ref{eq:Ez-1}) contain only
one mode (polaritonic one with index $n=0$) propagating towards the
interface in PC1 (referring to the above-mentioned incident wave
with amplitude ${\cal H}_{1,+}^{(0)}$) and a full set of modes propagating
backward from the interface (corresponding to the reflected harmonics
with amplitudes ${\cal H}_{1,-}^{(n)}$). At the same time, Eqs.(\ref{eq:Hy-2})
and (\ref{eq:Ez-2}) contain only modes propagating in PC2
in the positive direction of $x$-axis, which correspond to the transmitted
modes with amplitudes ${\cal H}_{2,+}^{(m)}$.

The next step is to apply the boundary conditions at the interface
$x=0$ {[}continuity of tangential components of magnetic field $H_{1,y}(0,z)=H_{2,y}(0,z)$,
and electric field $E_{1,z}(0,z)=E_{2,z}(0,z)${]} and use the
orthogonality of the spatial profile functions,
\[
\int_{md}^{md+d}\psi_{j}^{(n^{\prime})}\left(z\right)\overline{\psi_{j}^{(n)}}\left(z\right)dz=\delta_{n,n^{\prime}}d,
\]
where the overbar denotes complex conjugation.
After applying this orthogonality conditions to
Eqs.(\ref{eq:Hy-1})--(\ref{eq:Ez-2}) we obtain
the following equations for the amplitudes:
\begin{eqnarray}
\delta_{n,0}{\cal H}_{1,+}^{(0)}+{\cal H}_{1,-}^{(n)}=\sum_{l=0}^\infty{\cal H}_{2,+}^{(l)}\Psi_{l,n},\label{eq:step-final-1}\\
\left[\delta_{n,0}{\cal H}_{1,+}^{(0)}-{\cal H}_{1,-}^{(n)}\right]k_{1,x}^{(n)}=\sum_{l=0}^\infty k_{2,x}^{(l)}{\cal H}_{2,+}^{(l)}\Psi_{l,n},\label{eq:step-final-2}
\end{eqnarray}
where
\begin{equation}
\Psi_{l,n}=\frac{1}{d}\int_{md}^{md+d}\psi_{2}^{(l)}\left(z\right)\overline{\psi_{1}^{(n)}}\left(z\right)dz.\label{eq:coup-int}
\end{equation}
It should be noticed that the obtained Eqs.(\ref{eq:step-final-1})
and (\ref{eq:step-final-2}) can also be applied to the case of an
interface between the PC and a homogeneous medium (formally in this
case $E_{F2}=0$). In this case the spatial profile functions will
be as follows:
\begin{eqnarray}
\psi_{2}^{(l)}\left(z\right)=\exp\left(ik_{z}^{(l)}z\right),\label{eq:eigen-hd-psi}\\
k_{z}^{(l)}=q+\frac{2l}{d}\pi=\sqrt{\kappa^{2}\varepsilon-\left(k_{x}^{(l)}\right)^{2}}.\label{eq:eigen-hd-kz}
\end{eqnarray}
In order to express the reflection and transmission coefficients in terms of energy
fluxes, we notice that the component of the Poynting vector along the
direction of propagation ($x$-axis) for the $n$-th mode
$S_{j,\pm}^{(n)}=-\left(c/8\pi\right)\mathrm{Re}\left(E_{j,z,\pm}^{(n)}\overline{H_{j,y,\pm}^{(n)}}\right)$,
after substututing the explicit forms of the electromagnetic fields
(\ref{eq:Hy-1})--(\ref{eq:Ez-2}), can be written as
\begin{eqnarray*}
S_{j,\pm}^{(n)}=\pm\frac{c}{8\pi\kappa\varepsilon}\mathrm{Re}\left(k_{j,x}^{(n)}\right)\left|{\cal H}_{j,\pm}^{(n)}\right|^{2}\left|\psi_{j}^{(n)}\left(z\right)\right|^{2}.
\end{eqnarray*}
In other words, if the mode is propagating (with purely real $k_{x}^{(n)}$),
it carries energy either in positive (sign ``+'') or in negative
(sign ``-'') direction of $x$-axis. In contrast, evanescent modes (with
purely imaginary $k_{x}^{(n)}$) do not carry any energy. Thus,
we define the coefficients $R_{n}$, $T_{l}$ as the integral characteristics
\begin{eqnarray}
R_{n}=-\frac{\int_{md}^{md+d}S_{1,-}^{(n)}dz}{\int_{md}^{md+d}S_{1,+}^{(0)}dz}=\frac{\mathrm{Re}\left(k_{1,x}^{(n)}\right)\left|{\cal H}_{1,-}^{(n)}\right|^{2}}{\mathrm{Re}\left(k_{1,x}^{(0)}\right)\left|{\cal H}_{1,+}^{(0)}\right|^{2}},\label{eq:Rn-step}\\
T_{l}=\frac{\int_{md}^{md+d}S_{2,+}^{(l)}dz}{\int_{md}^{md+d}S_{1,+}^{(0)}dz}=\frac{\mathrm{Re}\left(k_{2,x}^{(l)}\right)\left|{\cal H}_{2,+}^{(l)}\right|^{2}}{\mathrm{Re}\left(k_{1,x}^{(0)}\right)\left|{\cal H}_{1,+}^{(0)}\right|^{2}}.\label{eq:Tn-step}
\end{eqnarray}
The coefficients $R_{0}$, $T_{0}$ are the reflectance and the transmittance
of the polaritonic mode, respectively, while the others ($n\text{\ensuremath{\ne}}0$)
are normalized intensities of higher diffraction orders in PC1 and
PC2 (coefficients $R_{n}$ and $T_{l}$, respectively).

\begin{figure*}
\includegraphics[height=8cm]{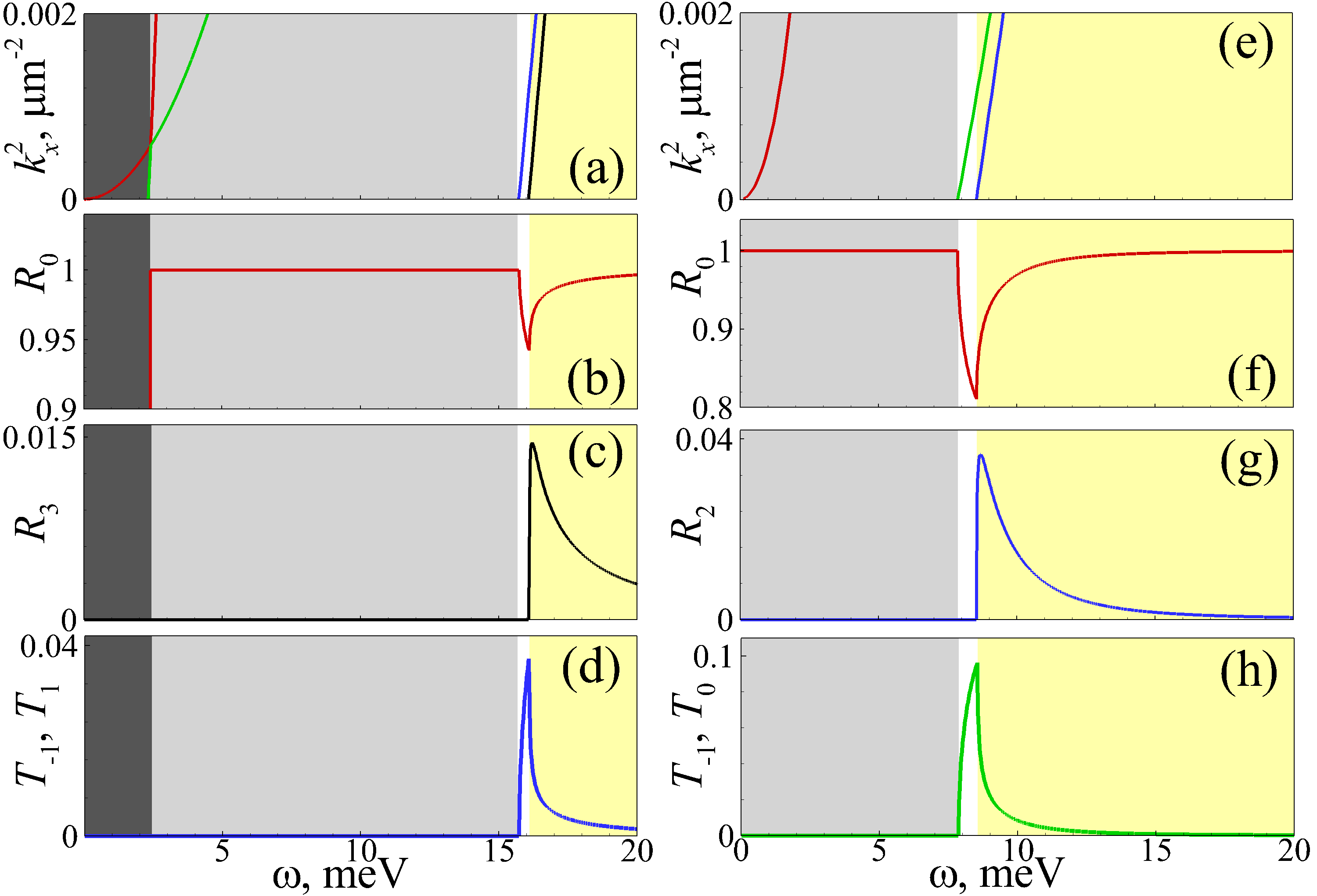}\includegraphics[height=8cm]{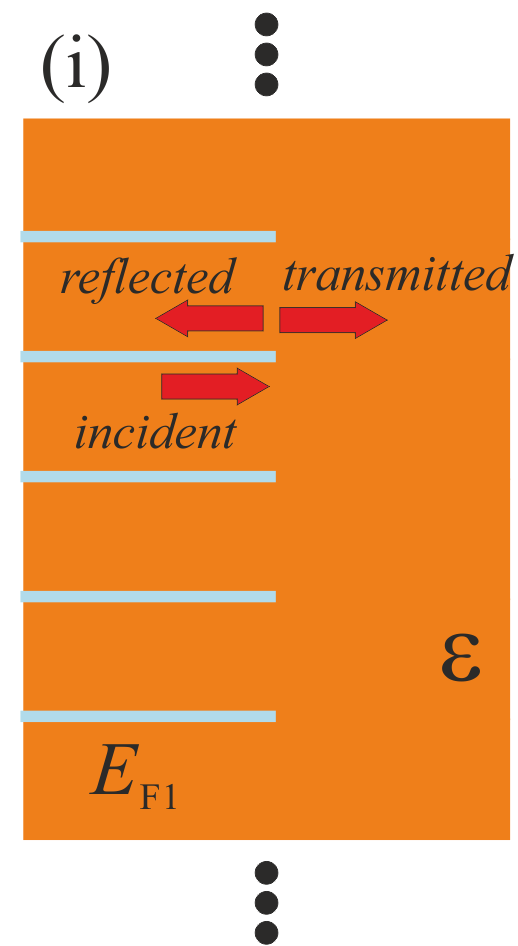}

\caption{(a,e) Dispersion curves {[}the same as in Fig.\ref{fig:sl-p}(a){]}
of graphene multilayer PC for $q=0$ {[}panel (a){]}, or for $q=\pi/d$ {[}panel (e){]}; (b)--(d), (f)--(h)
Frequency dependence of the reflectance $R_{n}\left(\omega\right)$
of polaritonic mode $n=0$ {[}panels (b) and (f){]}, diffraction order
intensities of PC modes $n=3$ or $n=2$ {[}correspondingly panels
(c) and (g){]} and those $T_{l}\left(\omega\right)$ of homogeneous
dielectric modes $l=-1$ and $l=-1$ {[}panel (d){]} or $l=-1$ and
$l=0$ {[}panel (h){]} for the case when the polaritonic mode with
Bloch wavevector $q=0$ {[}left column, panels (b)--(d){]} or $q=\pi/d$
{[}central column, panels (f)--(h){]} is scattered from the interface
between the graphene multilayer PC and homogeneous medium. The parameters
of the structure under consideration are $\varepsilon=3.9$, $E_{F1}=0.157\,$eV,
$d=40\,\mu\mathrm{m}$. Notice that in panel (d) transmission coefficients
for modes $l=-1$ and $l=1$ are equal to each other {[}as well as
in panel (h) transmission coefficients for modes for modes $l=-1$
and $l=0$ are also equal{]}; (i) Schematic view of the interface
between the graphene-based PC and a homogeneous medium. \label{fig:step-vac}}
\end{figure*}
First, we will consider the scattering of an incident polaritonic mode
on the interface between the graphene multilayer PC and a homogeneous
dielectric, schematically depicted in Fig.\ref{fig:step-vac}(i).
As it was mentioned, in the case $q=0$ {[}left column in Fig.\ref{fig:step-vac}{]}
the polaritonic mode exists in the frequency range $\omega>\omega_{*1}\approx2.4\,$meV
only {[}light gray, white and yellow domains in Figs.\ref{fig:step-vac}(a)--\ref{fig:step-vac}(d){]},
while the existence of polaritonic mode below the cutoff frequency, $\omega<\omega_{*1}$
is impossible {[}dark gray domains in in Figs.\ref{fig:step-vac}(a)--\ref{fig:step-vac}(d){]}.
In the frequency range $\omega_{*1}<\omega<2\pi c/\sqrt{\varepsilon}d\approx15.8\,$meV
{[}light gray domain in Figs.\ref{fig:step-vac}(a)--\ref{fig:step-vac}(d){]}
there are only two propagating modes in the PC1, namely, polaritonic
mode with $n=0$ {[}depicted by red line in Fig.\ref{fig:step-vac}(a){]}
and the light-line mode $n=1$ {[}depicted by green line in Fig.\ref{fig:step-vac}(a){]}.
In the homogeneous dielectric in this frequency range there is only
one propagating mode with $l=0$ {[}see Eq.(\ref{eq:eigen-hd-psi}){]}.
Notice the coincidence of both the shape and dispersion properties of
modes with $l=0$ in the homogeneous dielectric and the light-line mode
in PC1. Nevertheless, due to the opposite parity\cite{Note1} of the spatial profile
functions integral (\ref{eq:coup-int})$\Psi_{0,0}\left(0\right)\equiv0$,
i.e. polaritonic mode can not be coupled to the light-line mode,
thus giving rise to the total reflection ($R_{0}=1$) of the polaritonic
mode in this frequency range, shown in Fig.\ref{fig:step-vac}(b).

The narrow frequency range $15.8\,\mathrm{meV}\lesssim\omega\lesssim16.1\,\mathrm{meV}$
{[}white domain in Figs.\ref{fig:step-vac}(a)--\ref{fig:step-vac}(d){]}
corresponds to the situation when one more mode in PC with $n=2$
{[}depicted by blue line in Fig.\ref{fig:step-vac}(a){]} as well
as two more modes in the homogeneous dielectric, with $l=-1,1$ become
propagating. In spite of having the same dispersion properties {[}compare
Eqs.(\ref{eq:w-g-0}) and (\ref{eq:eigen-hd-kz}){]}, these modes possess
different spatial profiles {[}compare Eqs.(\ref{eq:shape-w-0}) and
(\ref{eq:eigen-hd-psi}){]}. The incident polaritonic mode $n=0$,
although not being able to couple to the mode $n=2$ in PC due to the
opposite parity {[}see modes B in Fig.\ref{fig:sl-p}(b) and D in
Fig.\ref{fig:sl-p}(d){]}, still can couple to the modes
$l=-1,1$ in the homogeneous dielectric. The last fact results in
the nonzero intensities of these diffraction orders $T_{-1}$ and
$T_{1}$, shown in Fig.\ref{fig:step-vac}(d), and a decrease of the
polaritonic mode reflectance {[}Fig.\ref{fig:step-vac}(b){]}. At
higher frequencies, $\omega\gtrsim16.1\,$meV {[}yellow domain in Figs.\ref{fig:step-vac}(a)--\ref{fig:step-vac}(d){]}
there is one more propagating mode with $n=3$ in the PC1 with the same
parity as the incident polaritonic mode {[}compare modes B in Fig.\ref{fig:sl-p}(b)
and E in Fig.\ref{fig:sl-p}(e){]}, thus allowing their coupling
and nonzero intensity of diffraction order $R_{3}$ {[}Fig.\ref{fig:step-vac}(c){]}.
Nevertheless, the third diffraction order intensity $R_{3}$ is a decreasing
function of frequency. At the same time, the diffraction intensities
$T_{-1},\,T_{1}$ and the reflectance $R_{0}$ {[}see Figs.\ref{fig:step-vac}(d)
and \ref{fig:step-vac}(b){]} possess a maximum and a minimum, correspondingly,
at the boundary of the white and the yellow domains, 
$\omega\approx16.1\,$meV.

The diffraction of the polaritonic mode with $q=\pi/d$ {[}central
column in Fig.\ref{fig:step-vac}{]} is characterized by the following interesting features. In this case the polaritonic mode $n=0$ can exist
at any frequency {[}red line in Figs.\ref{fig:step-vac}(e){]}, thus
total reflectance $(R_{0}\equiv1)$ takes place {[}Fig.\ref{fig:step-vac}(f){]}
in the frequency range $\omega\lesssim7.9\,$meV. This frequency range
is depicted by light gray shadow in Figs.\ref{fig:step-vac}(e)--\ref{fig:step-vac}(h).
In the frequency range $7.9\,\mathrm{meV}\lesssim\omega\lesssim8.5\,$meV
{[}white domain in Figs.\ref{fig:step-vac}(e)--\ref{fig:step-vac}(h){]}
the energy of the incident mode is partially transformed into that of
the propagating modes $l=-1,0$ of the homogeneous dielectric {[}green
line in Fig.\ref{fig:step-vac}(e){]}, giving rise to nonzero diffraction
orders intensities $T_{-1}$ and $T_{0}$ {[}Fig.\ref{fig:step-vac}(h){]}.
At the same time, in the frequency range $\omega\gtrsim8.5\,$meV {[}yellow
domain in Figs.\ref{fig:step-vac}(e)--\ref{fig:step-vac}(h){]} the
coupling between the polaritonic mode and the photonic mode with $n=2$
becomes possible {[}Fig.\ref{fig:step-vac}(g){]}.

\begin{figure}
\includegraphics[width=8.5cm]{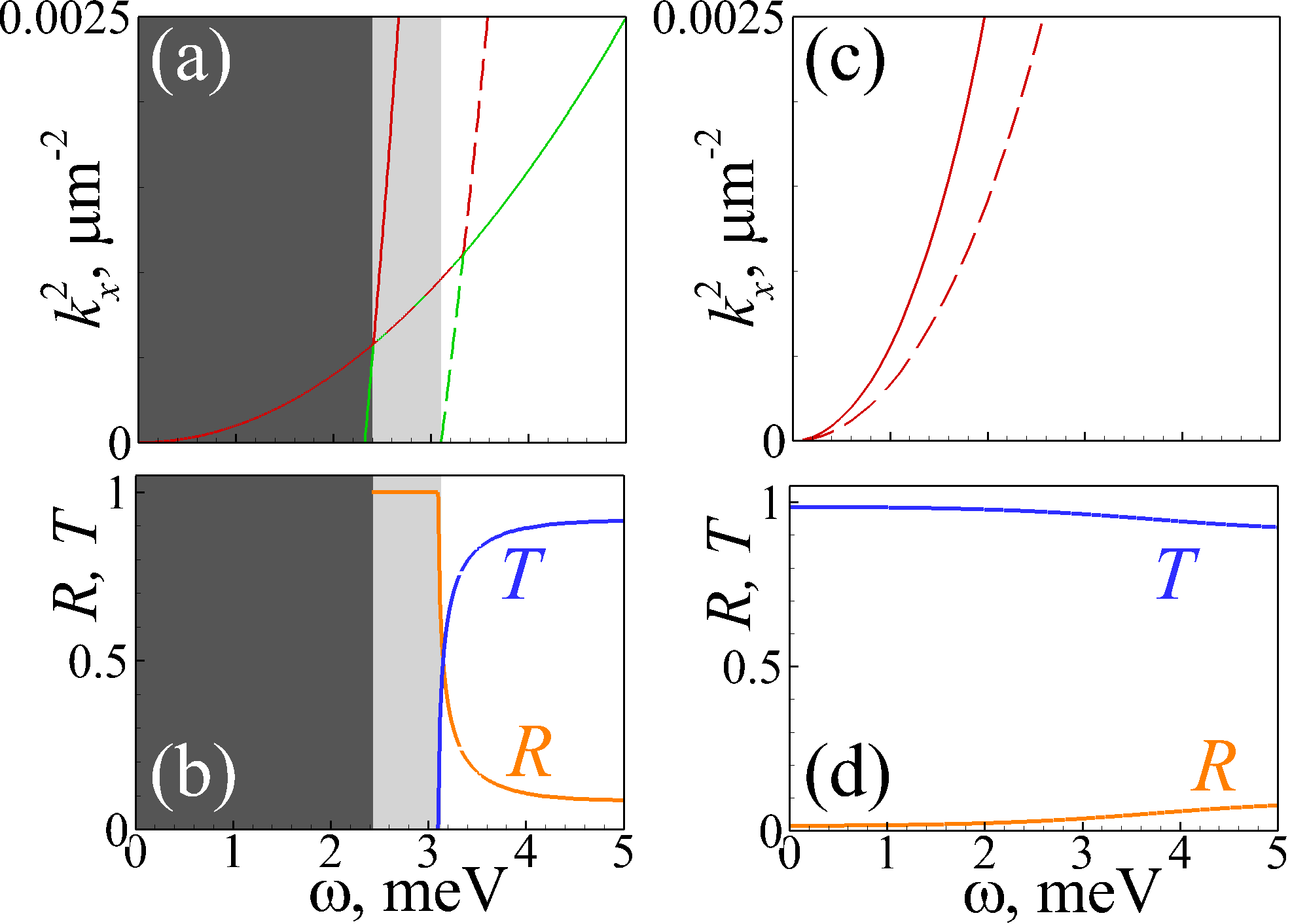}

\caption{(a), (c) Dispersion curves of graphene multilayer PC with $E_{F1}$
(solid lines) or $E_{F2}$ (dashed lines) for $q=0$ {[}panel
(a){]}, or for $q=\pi/d$ {[}panel (c){]}; (b),(d)
Frequency dependence of the reflectance $R_{0}\left(\omega\right)$
and the transmittance $T_{0}\left(\omega\right)$ of polaritonic mode
with Bloch wavevector $q=0$ {[}panel (b){]} or $q=\pi/d$ {[}panel
(d){]} when it is diffracted on the interface between two graphene
multilayer PCs with Fermi energies $E_{F1}$ and $E_{F2}$. The parameters
of the structure under consideration are: $\varepsilon=3.9$, $E_{F1}=0.157\,$eV,
$E_{F2}=0.3\, $eV, $d=40\,\mu\mathrm{m}$.\label{fig:step-lat}}
\end{figure}
The characteristics of the diffraction of the polaritonic mode at
the interface between two graphene multilayer PCs are shown in Fig.\ref{fig:step-lat}.
As in the previous case, polaritonic modes for $q=0$ exist above
the cutoff frequencies $\omega_{*1}$ and $\omega_{*2}$ in the PC1 and PC2, respectively {[}nonexistence domain $\omega<\omega_{*1}$
is depicted in Figs.\ref{fig:step-lat}(a) and \ref{fig:step-lat}(b)
by dark gray color{]}. The frequency range $\omega_{*1}<\omega\lesssim3.1\,$meV
{[}light gray domain in Figs.\ref{fig:step-lat}(a) and \ref{fig:step-lat}(b){]}
is characterized by the existence of polaritonic mode in the PC1,
while in the PC2 there is only one propagating mode (the light-line
one). Owing to the above-mentioned opposite parity of these modes,
mutual coupling between them is not possible, which results in the
total reflection $R_{0}=1$ {[}see Fig.\ref{fig:step-lat}(b){]} of
the polaritonic mode from the interface between two PCs. In the frequency
range $\omega\gtrsim3.1\,$meV {[}white domain in Figs.\ref{fig:step-lat}(a)
and \ref{fig:step-lat}(b){]} the PC2 contains one more propagating
mode, which is photonic ($n=1$) below the cutoff frequency $\omega_{*2}$
and polaritonic ($n=0$) above it. The last fact gives rise to
a gradual decrease of the reflectance and an increase of the transmittance
{[}Fig.\ref{fig:step-lat}(b){]} of the incident polaritonic mode. At
the same time, when $q=\pi/d$ there is no cutoff frequency for the
existence of polaritonic mode {[}see Fig.\ref{fig:step-lat}(c){]},
which results in the nonzero transmittance at any frequency {[}Fig.\ref{fig:step-lat}(d){]}.
It is interesting that for $q=\pi/d$ (contrary to the
case of $q=0$) the reflectance (transmittance) is an increasing (decreasing)
function of frequency {[}compare Figs.\ref{fig:step-lat}(b) and \ref{fig:step-lat}(d){]}.

\section{Scattering of polaritonic mode from the double interface between
two graphene multilayer PCs}

\label{sec:double-interface}We now consider the situation {[}Fig.\ref{fig:geometry_step_barrier}(b){]}
when a graphene multilayer PC of finite width $D$ (along the $x$-axis)
and with graphene layer's Fermi energy $E_{F2}$ (further referred to as PC2) is cladded by
two semi-infinite PCs which graphene layer are characterized by the
Fermi energy $E_{F1}$ (these two PCs will be referred to as PC1 and PC3). As in Sec.\ref{sec:eigen}, the incident wave
is the polaritonic mode of PC1, propagating in the positive direction
of the $x$-axis. Notice, that the only difference between Fig.\ref{fig:geometry_step_barrier}(a)
and \ref{fig:geometry_step_barrier}(b) is the presence of two
interfaces in the last case.

We note that the electromagnetic field in the region $x<0$ can be
represented in the same manner as in Eqs.(\ref{eq:Hy-1}) and (\ref{eq:Ez-1}),
while inside the region $0<x<D$, occupied by the PC2, the field components will have the form:
\begin{eqnarray}
H_{2,y}(x,z)=\sum_{l}\psi_{2}^{(l)}\left(z\right)\times\nonumber \\
\left[{\cal H}_{2,+}^{(l)}\exp(ik_{2,x}^{(l)}x)+{\cal H}_{2,-}^{(l)}\exp(-ik_{2,x}^{(l)}x)\right],\label{eq:Hy-2-bar}\\
E_{2,z}(x,z)=-\sum_{l}\frac{k_{2,x}^{(l)}}{\kappa\varepsilon}\psi_{2}^{(l)}\left(z\right)\times\nonumber \\
\left[{\cal H}_{2,+}^{(l)}\exp(ik_{2,x}^{(l)}x)-{\cal H}_{2,-}^{(l)}\exp(-ik_{2,x}^{(l)}x)\right].\label{eq:Ez-2-bar}
\end{eqnarray}
Due to the finite width $D$ of the PC2, Eqs.(\ref{eq:Hy-2-bar})
and (\ref{eq:Ez-2-bar}) contain both forward- and backward propagating waves. Finally, in the PC3 (region $x>D$) the electromagnetic field
components are:
\begin{eqnarray}
H_{3,y}(x,z)=\sum_{n}{\cal H}_{3,+}^{(n)}\psi_{1}^{(n)}\left(z\right)\times\nonumber \\
\exp\left[ik_{1,x}^{(n)}(x-D)\right],\label{eq:Hy-3-bar}\\
E_{3,z}(x,z)=-\sum_{n}{\cal H}_{3,+}^{(n)}\frac{k_{1,x}^{(n)}}{\kappa\varepsilon}\psi_{1}^{(n)}\left(z\right)\times\nonumber \\
\exp\left[ik_{1,x}^{(n)}(x-D)\right].\label{eq:Ez-3-bar}
\end{eqnarray}
\begin{figure}
\includegraphics[width=8.5cm]{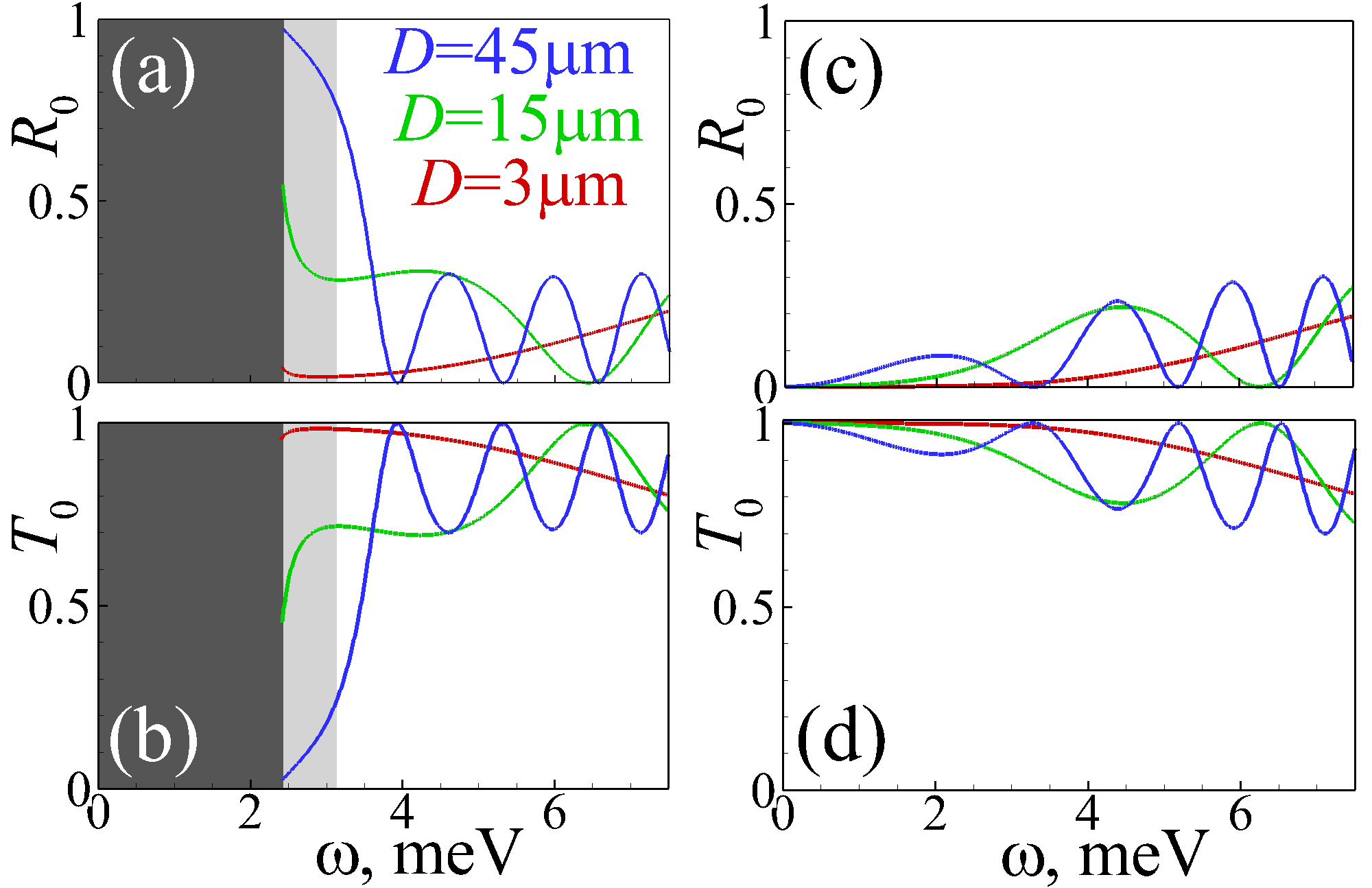}

\caption{Reflectance {[}panels (a) and (c){]} and transmittance {[}panels (b)
and (d){]} versus frequency $\omega$ for the double interface between
two graphene multilayer PCs with parameters $\varepsilon=3.9$, $E_{F1}=0.157\,$eV,
$E_{F2}=0.3\,$eV,$d=40\,\mu\mathrm{m}$, $q=0$ {[}panels (a) and
(b){]}, or $q=\pi/d$ {[}panels (c) and (d){]}, $D=45\,\mu$m (blue
lines), $D=15\,\mu$m (green lines), $D=3\,\mu$m (red lines).\label{fig:barrier-lat}}
\end{figure}
Applying the same boundary condition as in Sec.\ref{sec:single-interface},
we have:
\begin{eqnarray}
\delta_{n,0}{\cal H}_{1,+}^{(0)}+{\cal H}_{1,-}^{(n)}=\sum_{l}\Psi_{l,n}\left[{\cal H}_{2,+}^{(l)}+{\cal H}_{2,-}^{(l)}\right],\label{eq:bar-final-1}\\
\left[\delta_{n,0}{\cal H}_{1,+}^{(0)}-{\cal H}_{1,-}^{(n)}\right]k_{1,x}^{(n)}=\label{eq:bar-final-2}\\
\sum_{l}k_{2,x}^{(l)}\Psi_{l,n}\left[{\cal H}_{2,+}^{(l)}-{\cal H}_{2,-}^{(l)}\right],\nonumber \\
{\cal H}_{3,+}^{(n)}=\sum_{l}\Psi_{l,n}\times\label{eq:bar-final-3}\\
\left[{\cal H}_{2,+}^{(l)}\exp(ik_{2,x}^{(l)}D)+{\cal H}_{2,-}^{(l)}\exp(-ik_{2,x}^{(l)}D)\right],\nonumber \\
{\cal H}_{3,+}^{(n)}k_{1,x}^{(n)}=\sum_{l}k_{2,x}^{(l)}\Psi_{l,n}\times\label{eq:bar-final-4}\\
\left[{\cal H}_{2,+}^{(l)}\exp(ik_{2,x}^{(l)}D)-{\cal H}_{2,-}^{(l)}\exp(-ik_{2,x}^{(l)}D)\right].\nonumber
\end{eqnarray}
For this structure the coefficients $T_{n}$ are expressed as
\begin{eqnarray*}
T_{n}=\frac{\int_{md}^{md+d}S_{2,+}^{(n)}dz}{\int_{md}^{md+d}S_{1,+}^{(0)}dz}=\frac{\mathrm{Re}\left(k_{1,x}^{(n)}\right)\left|{\cal H}_{3,+}^{(n)}\left(q\right)\right|^{2}}{\mathrm{Re}\left(k_{1,x}^{(0)}\right)\left|{\cal H}_{1,+}^{(0)}\left(q\right)\right|^{2}},
\end{eqnarray*}
 while the coefficients $R_{n}$ are the same as before [Eq.(\ref{eq:Rn-step})].

The frequency dependence of the reflectance and transmittance for this
structure are shown in Fig.\ref{fig:barrier-lat}. For $q=0$ and
large width of the PC2 {[}light gray domain in Figs.\ref{fig:barrier-lat}(a) and \ref{fig:barrier-lat}(b){]}
the reflectance {[}blue line in Fig.\ref{fig:barrier-lat}(a){]} of
the structure is slightly less than unity in the frequency range $\omega_{*1}<\omega\lesssim3.1\,$meV and it is a decreasing function
of the frequency, while the transmittance is nonzero and is an increasing
function of $\omega $ (compare with Fig.\ref{fig:step-lat}(b), where
in the case of single interface this range corresponds to the total
reflectance $R_{0}=1$, $T_{0}=0$). The reason for this is the tunneling
through the region $0<x<D$ because there is no propagating mode in PC2. The tunneling rate (and, as a consequence, the transmittance $T_{0}$)
gradually grows when the PC2 width, $D$ is decreased, compare
blue, green and red lines in Fig.\ref{fig:barrier-lat}(b). Notice
that for small $D=3\,\mu$m {[}red line in Fig.\ref{fig:barrier-lat}(b){]}
almost the whole energy of the incident wave is transmitted via tunneling, thus giving $T_{0}\lesssim1$.
In the frequency range $\omega\gtrapprox3.1\,$meV {[}white domain in
Figs.\ref{fig:barrier-lat}(a) and \ref{fig:barrier-lat}(b){]} one
more mode in PC2 becomes propagating, which gives rise to the
Fabry-P\'erot oscillations of the transmittance and reflectance and the
possibility of total transmission $T_{0}=1$ (respectively, $R_{0}=0$).
The latter takes place at frequencies, for which the width $D$ matches an integer number of polaritonic
mode half-wavelengths, that is
\begin{equation}
k_{2,x}^{(0)}D=l\pi,\qquad l\ge1.\label{eq:FP}
\end{equation}
The same phenomenon also occurs when $q=\pi/d$ {[}see Figs.\ref{fig:barrier-lat}(c)
and \ref{fig:barrier-lat}(d){]}. The frequencies at which full transmission
takes place can be approximated using the Drude model for graphene's conductivity.
Correspondingly, the parameter $\Lambda$ in the dispersion relation
(\ref{eq:disp-p}) can be approximated by $\Lambda=2\alpha cE_{F2}/\hbar\omega^{2}\varepsilon$,
while we also use the so-called non-retarded approximation for $k_{2,x}^{(0)}=-ik_{z}^{(0)}$.
Under these assumptions the dispersion relation (\ref{eq:disp-p})
can be rewritten as
\[
\cos\left(qd\right)-\cosh\left(k_{2,x}^{(0)}d\right)+\Lambda k_{2,x}^{(0)}\sinh\left(k_{2,x}^{(0)}d\right)=0,
\]
 which along with Eq.(\ref{eq:FP}) gives
\begin{equation}
\omega^{2}=\frac{2\alpha cE_{F2}}{\hbar\varepsilon}\frac{l\pi}{D}\frac{\sinh\left(\frac{l\pi}{D}d\right)}{\cosh\left(\frac{l\pi}{D}d\right)-\cos\left(qd\right)}.\label{eq:FP-appr}
\end{equation}
For the case $D=15\mu$m Eq.(\ref{eq:FP-appr}) gives $\omega\approx6.82\,$meV
for $q=0$ and $\omega\approx6.81\,$meV for $q=\pi/d$, which qualitatively
agrees with Figs.\ref{fig:barrier-lat}(b) and \ref{fig:barrier-lat}(d)
{[}green line{]}.

The oscillations of the transmittance can be seen as an optical analogue of the well known effect of electron resonant tunneling in a double-barrier heterostructure (DBHS).\cite{Esaki} The modulation depth of the electromagnetic wave transmission is smaller than in the case of electrons because of the weaker confinement of the former (so that direct tunneling involving only evanescent waves in the PC2 is possible) and also because of the simultaneous presence of several scattering channels.

\section{Conclusions}

We have analysed the scattering of surface plasmon-polaritons generated in a graphene multilayer photonic crystal and propagating across its lateral surface or interface with another graphene multilayer PC with a different band structure (controlled by the graphene's Fermi energy).  In particular, for the low-frequency region where only the polaritonic mode (with imaginary $z$ component of the wavevector, $k_z$) is propagating in the direction along the graphene sheets (while the modes with real $k_z$ have imaginary $k_x$, i.e. are evanescent in this sense), we have shown that this mode is totally reflected from the interface between the graphene
multilayer PC and a homogeneous dielectric. Nevertheless, in the higher-frequency
region where photonic (i.e. propagating with real $k_z$) modes are allowed, the partial
transformation of the incident polaritonic mode's energy into that of other
diffraction orders (both in PC and in homogeneous dielectric) becomes
possible, thus reducing the reflectance of the incident wave. Moreover,
by virtue of the reciprocity principle this gives rise to the possibility to
excite the PC's polaritonic eigenmode by an external wave impinging on its
edge. In-phase polaritonic mode (with $q=0$) can also be totally
reflected from the interface between two PCs, while for out-of-phase
oscillations (with $q\ne0$) this scenario is impossible. It is also shown
that the transmittance and the reflectance of a structure consisting of three photonic crystals and two interfaces between
them (we asumed PC1=PC3$\neq $PC2) exhibit Fabry-P\'erot oscillations with a discrete set of frequencies for which total transmission of the polaritonic mode takes place.
This effect is similar to the resonant tunneling of electrons in a DBHS where the electric current shows a sharp peak as a function of bias.\cite{Esaki} 
Here, in addition to the difference between the Fermi levels of the crystals PC1 and PC2, the electromagnetic wave transmission depends also on the PC wavevector $q$.

\section*{Acknowledgments}
We acknowledge support from the EC under the
Graphene Flagship (Contract No. CNECT-ICT-649953).

\appendix

\section{Approximations for the eigenvalues and eigenfunctions.}

Let us consider first the asymptotical behaviour of the polaritonic
modes $n=0$ at high frequencies, where the modes with different Bloch
wavevector $q$ merge. Since SPPs are evanescent
waves, it is natural to introduce a decay parameter in $z$-direction,
$p=-ik_{z}^{(0)}$. In this case the dispersion relation (\ref{eq:disp-p})
can be rewritten as
\begin{equation}
p=\frac{\cosh\left(pd\right)-\cos\left(qd\right)}{\Lambda\sinh\left(pd\right)}.\label{eq:disp-p-forp}
\end{equation}
Notice that in the limiting case $d\to\infty$ the dispersion relation
(\ref{eq:disp-p-forp}) transforms into $p=\Lambda^{-1}$, i.e. into
the dispersion relation for a single graphene sheet. If we consider
the situation of $d$ being large but finite, the single sheet dispersion
relation can be used as zeroth approximation for $p$. This approximation after being sibstituted into the right hand side of Eq.(\ref{eq:disp-p-forp}), allows for an approximate respresentation of Eq.(\ref{eq:disp-p-forp}) as
\begin{equation}
p=\sqrt{\left(k_{x}^{(0)}\right)^{2}-\kappa^{2}\varepsilon}=\frac{1-\cos\left(qd\right)\exp\left(-d/\Lambda\right)}{\Lambda}.\label{eq:disp-p-forp-appr}
\end{equation}
The above-mentioned zeroth approximation for $p$ also can be used
for approximating the spatial profile function (\ref{eq:psi}), namely:
\begin{eqnarray*}
\psi^{(0)}\left(z||q,\omega\right)=\left\{ \exp\left[iqd\right]\cosh\left[\Lambda^{-1}\left(z-md\right)\right]-\right.\\
\left.\cosh\left[\Lambda^{-1}\left(md+d-z\right)\right]\right\} \frac{\exp\left[iqmd\right]}{\sqrt{A^{(0)}}},\\
A^{(0)}=1-\cos\left(qd\right)\cosh\left(\Lambda^{-1}d\right)+\\
\Lambda\frac{\cosh\left(\Lambda^{-1}d\right)-\cos\left(qd\right)}{d}\sinh\left(\Lambda^{-1}d\right).
\end{eqnarray*}

Even more, since all the phenomena we are interested in take place in
the frequency range $\hbar\omega\ll E_{F}$, we can take into account only
the Drude contribution to graphene's conductivity.
Correspondingly, the parameter $\Lambda$ can be approximated by
$\Lambda=2\alpha cE_{F}/\hbar\omega^{2}\varepsilon$.

This fact can be used in the approximation of the touching point (cutoff
frequency) between the zeroth and first band at $q=0$, mentioned in
Sec.\ref{sec:eigen}. Taking into acccount the smallness of $k_{z}$ in the 
touch points, one can expand the dispersion relation (\ref{eq:disp-p}) as
\[
\cos\left(qd\right)-1+\frac{\left(k_{z}d\right)^{2}}{2}-\frac{\left(k_{z}d\right)^{4}}{24}-\Lambda k_{z}\left[k_{z}d-\frac{\left(k_{z}d\right)^{3}}{6}\right]=0.
\]
At $q=0$ we have
\[
k_{z}^{2}=\frac{12}{d^{2}}\frac{d-2\Lambda}{d-4\Lambda}.
\]
Taking into account the above-mentioned approximation for $\Lambda$
we find that $k_{z}^{2}>0$ (with $k_{z}^{(0)}$ being purely real)
or $k_{z}^{2}<0$ (with $k_{z}^{(1)}$ being purely imaginary) in the
frequency ranges $\omega<\omega_{*}=\left(4\alpha E_{F}c/\hbar d\varepsilon\right)^{1/2}$
and $\omega>\omega_{*}$, respectively. The last fact well corroborates
with the dispersion curve behavior in the vicinity of the cutoff frequency $\omega_{*}$,
depicted in Fig.\ref{fig:sl-p}(a).

From Fig.\ref{fig:sl-p}(a) it is evident that in the high-frequency
region $\omega\gg\omega_{*}$ the gaps becomes narrow. We can use this
fact in order to approximate the eigenfunctions and the eigenvalues in this
region. At the edge of the Brillouin zone ($q=\pi/d$) we represent
$z$-component of the wavevector in the band with even number $n=2l$
($1\le l<\infty$) as $k_{z}^{(2l)}=k_{z}^{(2l-1)}+\Delta^{(2l)}$
with $\Delta^{(2l)}$ being small value ($\left|\Delta^{(2l)}\right|\ll\left|k_{z}^{(2l-1)}\right|$)
and $k_{z}^{(2l-1)}$ is determined in Eq.(\ref{eq:w-g-pi}). In this
case the dispersion relation after the expansion of sine and cosine
with respect to $\Delta^{(2l)}$can be represented as
\[
-\frac{\left(\Delta^{(2l)}d\right)^{2}}{2}+\Lambda\left[\frac{2l-1}{d}\pi+\Delta^{(2l)}\right]\Delta^{(2l)}d=0,
\]
 from which it is possible to determine
\[
\Delta^{(2l)}=\frac{2l-1}{d}\pi\frac{2\Lambda}{d-2\Lambda}
\]
 and
\[
k_{z}^{(2l)}=\pi\frac{2l-1}{d-2\Lambda}.
\]
In the same manner it is possible to obtain an approximate expression
for the spatial profile function (\ref{eq:psi}):
\begin{eqnarray}
\psi^{(2l)}\left(z||\frac{\pi}{d},\omega\right)=\left(-1\right)^{m}\left\{ \sin\left[\frac{2l-1}{d}\pi z\right]-\right.\nonumber \\
\frac{\Delta^{(2l)}}{2}\cos\left[\frac{2l-1}{d}\pi z\right]\left[\left(2m+1\right)d-2z\right]-\label{eq:psi-pi}\\
\frac{\left(\Delta^{(2l)}\right)^{2}}{6}\sin\left[\frac{2l-1}{d}\pi z\right]\times\nonumber \\
\left.\left[d^{2}-3\left(z-md\right)\left(z-md-d\right)\right]\right\} \sqrt{\frac{2}{A^{(2l)}}},\nonumber
\end{eqnarray}
with
\begin{eqnarray*}
A^{(2l)}=1-\frac{\Delta^{(2l)}d}{\left(2l-1\right)\pi}-\\
\left(\frac{1}{12}-\frac{1}{\left(2l-1\right)^{2}\pi^{2}}\right)\left(\Delta^{(2l)}d\right)^{2}.
\end{eqnarray*}
The same formalism can be applied to the boundaries of the bands with an 
odd numbers $n=2l+1$ ($1\le l<\infty$) at the center of Brillouin
zone $q=0$. Thus, we obtain
\[
\Delta^{(2l+1)}=\frac{2l}{d}\pi\frac{2\Lambda}{d-2\Lambda}
\]
 and
\[
k_{z}^{(2l+1)}=k_{z}^{(2l)}+\Delta^{(2l+1)}=\pi\frac{2l}{d-2\Lambda}.
\]
An approximate expression for the spatial profile function (\ref{eq:psi})
can be written
\begin{eqnarray}
\psi^{(2l+1)}\left(z||0\right)=\left\{ \sin\left[\frac{2l}{d}\pi z\right]-\right.\nonumber \\
\frac{\Delta^{(2l+1)}}{2}\cos\left[\frac{2l}{d}\pi z\right]\left[\left(2m+1\right)d-2z\right]-\label{eq:psi-0}\\
\frac{\left(\Delta^{(2l+1)}\right)^{2}}{6}\sin\left[\frac{2l}{d}\pi z\right]\times\nonumber \\
\left.\left[d^{2}-3\left(z-md\right)\left(z-md-d\right)\right]\right\} \sqrt{\frac{2}{A^{(2l+1)}}},\nonumber
\end{eqnarray}
with
\begin{eqnarray*}
A^{(2l+1)}=1-\frac{\Delta^{(2l+1)}d}{2l\pi}-
\left(\frac{1}{12}-\frac{1}{4l^{2}\pi^{2}}\right)\left(\Delta^{(2l+1)}d\right)^{2}.
\end{eqnarray*}

\bibliographystyle{apsrev}
\bibliography{refl_sp_bib}

\begin{thebibliography}{43}
\expandafter\ifx\csname natexlab\endcsname\relax\def\natexlab#1{#1}\fi
\expandafter\ifx\csname bibnamefont\endcsname\relax
  \def\bibnamefont#1{#1}\fi
\expandafter\ifx\csname bibfnamefont\endcsname\relax
  \def\bibfnamefont#1{#1}\fi
\expandafter\ifx\csname citenamefont\endcsname\relax
  \def\citenamefont#1{#1}\fi
\expandafter\ifx\csname url\endcsname\relax
  \def\url#1{\texttt{#1}}\fi
\expandafter\ifx\csname urlprefix\endcsname\relax\def\urlprefix{URL }\fi
\providecommand{\bibinfo}[2]{#2}
\providecommand{\eprint}[2][]{\url{#2}}

\bibitem[{\citenamefont{Maier}(2007)}]{Maier}
\bibinfo{author}{\bibfnamefont{S.~A.} \bibnamefont{Maier}},
  \emph{\bibinfo{title}{Plasmonics: Fundamentals and Applications}}
  (\bibinfo{publisher}{Springer}, \bibinfo{address}{New York},
  \bibinfo{year}{2007}).

\bibitem[{\citenamefont{Low and Avouris}(2014)}]{ACSgp}
\bibinfo{author}{\bibfnamefont{T.}~\bibnamefont{Low}} \bibnamefont{and}
  \bibinfo{author}{\bibfnamefont{P.}~\bibnamefont{Avouris}},
  \bibinfo{journal}{ACS Nano} \textbf{\bibinfo{volume}{8}},
  \bibinfo{pages}{1086} (\bibinfo{year}{2014}).

\bibitem[{\citenamefont{Dionne and Atwater}(2012)}]{MRS:8669483}
\bibinfo{author}{\bibfnamefont{J.~A.} \bibnamefont{Dionne}} \bibnamefont{and}
  \bibinfo{author}{\bibfnamefont{H.~A.} \bibnamefont{Atwater}},
  \bibinfo{journal}{MRS Bulletin} \textbf{\bibinfo{volume}{37}},
  \bibinfo{pages}{717} (\bibinfo{year}{2012}).

\bibitem[{\citenamefont{Stockman}(2011)}]{Stockman}
\bibinfo{author}{\bibfnamefont{M.~I.} \bibnamefont{Stockman}},
  \bibinfo{journal}{Phys. Today} \textbf{\bibinfo{volume}{64}},
  \bibinfo{pages}{39} (\bibinfo{year}{2011}).

\bibitem[{\citenamefont{Khurgin and Boltasseva}(2012)}]{MRS:8669489}
\bibinfo{author}{\bibfnamefont{J.~B.} \bibnamefont{Khurgin}} \bibnamefont{and}
  \bibinfo{author}{\bibfnamefont{A.}~\bibnamefont{Boltasseva}},
  \bibinfo{journal}{MRS Bulletin} \textbf{\bibinfo{volume}{37}},
  \bibinfo{pages}{768} (\bibinfo{year}{2012}).

\bibitem[{\citenamefont{de~Abajo}(2014)}]{AbajoACSP}
\bibinfo{author}{\bibfnamefont{F.~J.~G.} \bibnamefont{de~Abajo}},
  \bibinfo{journal}{ACS Photonics} \textbf{\bibinfo{volume}{1}}
  (\bibinfo{year}{2014}).

\bibitem[{\citenamefont{Bludov et~al.}(2013{\natexlab{a}})\citenamefont{Bludov,
  Peres, and Vasilevskiy}}]{c:Bludov2013}
\bibinfo{author}{\bibfnamefont{Y.~V.} \bibnamefont{Bludov}},
  \bibinfo{author}{\bibfnamefont{N.~M.~R.} \bibnamefont{Peres}},
  \bibnamefont{and} \bibinfo{author}{\bibfnamefont{M.~I.}
  \bibnamefont{Vasilevskiy}}, \bibinfo{journal}{Journal of Optics}
  \textbf{\bibinfo{volume}{15}}, \bibinfo{pages}{114004}
  (\bibinfo{year}{2013}{\natexlab{a}}).

\bibitem[{\citenamefont{Stauber}(2014)}]{JPCM26}
\bibinfo{author}{\bibfnamefont{T.}~\bibnamefont{Stauber}}, \bibinfo{journal}{J.
  Phys.: Condens. Matter} \textbf{\bibinfo{volume}{26}},
  \bibinfo{pages}{123201} (\bibinfo{year}{2014}).

\bibitem[{\citenamefont{Chen et~al.}(2012)\citenamefont{Chen, Badioli,
  Alonso-Gonz\'alez, Thongrattanasiri, Huth, Osmond, Spasenovi\'c, Centeno,
  Pesquera, Godignon et~al.}}]{nanoImICFO}
\bibinfo{author}{\bibfnamefont{J.}~\bibnamefont{Chen}},
  \bibinfo{author}{\bibfnamefont{M.}~\bibnamefont{Badioli}},
  \bibinfo{author}{\bibfnamefont{P.}~\bibnamefont{Alonso-Gonz\'alez}},
  \bibinfo{author}{\bibfnamefont{S.}~\bibnamefont{Thongrattanasiri}},
  \bibinfo{author}{\bibfnamefont{F.}~\bibnamefont{Huth}},
  \bibinfo{author}{\bibfnamefont{J.}~\bibnamefont{Osmond}},
  \bibinfo{author}{\bibfnamefont{M.}~\bibnamefont{Spasenovi\'c}},
  \bibinfo{author}{\bibfnamefont{A.}~\bibnamefont{Centeno}},
  \bibinfo{author}{\bibfnamefont{A.}~\bibnamefont{Pesquera}},
  \bibinfo{author}{\bibfnamefont{P.}~\bibnamefont{Godignon}},
  \bibnamefont{et~al.}, \bibinfo{journal}{Nature}
  \textbf{\bibinfo{volume}{487}}, \bibinfo{pages}{77} (\bibinfo{year}{2012}).

\bibitem[{\citenamefont{Fei et~al.}(2012)\citenamefont{Fei, Rodin, Andreev,
  Bao, McLeod, Wagner, Zhang, Zhao, Thiemens, Dominguez et~al.}}]{nanoImBasov}
\bibinfo{author}{\bibfnamefont{Z.}~\bibnamefont{Fei}},
  \bibinfo{author}{\bibfnamefont{A.~S.} \bibnamefont{Rodin}},
  \bibinfo{author}{\bibfnamefont{G.~O.} \bibnamefont{Andreev}},
  \bibinfo{author}{\bibfnamefont{W.}~\bibnamefont{Bao}},
  \bibinfo{author}{\bibfnamefont{A.~S.} \bibnamefont{McLeod}},
  \bibinfo{author}{\bibfnamefont{M.}~\bibnamefont{Wagner}},
  \bibinfo{author}{\bibfnamefont{L.~M.} \bibnamefont{Zhang}},
  \bibinfo{author}{\bibfnamefont{Z.}~\bibnamefont{Zhao}},
  \bibinfo{author}{\bibfnamefont{M.}~\bibnamefont{Thiemens}},
  \bibinfo{author}{\bibfnamefont{G.}~\bibnamefont{Dominguez}},
  \bibnamefont{et~al.}, \bibinfo{journal}{Nature}
  \textbf{\bibinfo{volume}{487}}, \bibinfo{pages}{82} (\bibinfo{year}{2012}).

\bibitem[{\citenamefont{Gon\c{c}alves and Peres}(2016)}]{Andre}
\bibinfo{author}{\bibfnamefont{P.~A.~D.} \bibnamefont{Gon\c{c}alves}}
  \bibnamefont{and} \bibinfo{author}{\bibfnamefont{N.~M.~R.}
  \bibnamefont{Peres}}, \emph{\bibinfo{title}{An Introduction to Graphene
  Plasmonics}} (\bibinfo{publisher}{World Scientific},
  \bibinfo{address}{Singapore}, \bibinfo{year}{2016}).

\bibitem[{\citenamefont{Koppens et~al.}(2011)\citenamefont{Koppens, Chang, and
  de~Abajo}}]{nlgp}
\bibinfo{author}{\bibfnamefont{F.~H.~L.} \bibnamefont{Koppens}},
  \bibinfo{author}{\bibfnamefont{D.~E.} \bibnamefont{Chang}}, \bibnamefont{and}
  \bibinfo{author}{\bibfnamefont{F.~J.~G.} \bibnamefont{de~Abajo}},
  \bibinfo{journal}{Nano Lett.} \textbf{\bibinfo{volume}{11}},
  \bibinfo{pages}{3370 } (\bibinfo{year}{2011}).

\bibitem[{\citenamefont{Berman et~al.}(2010)\citenamefont{Berman, Boyko,
  Kezerashvili, Kolesnikov, and Lozovik}}]{c:Berman2010}
\bibinfo{author}{\bibfnamefont{O.~L.} \bibnamefont{Berman}},
  \bibinfo{author}{\bibfnamefont{V.~S.} \bibnamefont{Boyko}},
  \bibinfo{author}{\bibfnamefont{R.~Y.} \bibnamefont{Kezerashvili}},
  \bibinfo{author}{\bibfnamefont{A.~A.} \bibnamefont{Kolesnikov}},
  \bibnamefont{and} \bibinfo{author}{\bibfnamefont{Y.~E.}
  \bibnamefont{Lozovik}}, \bibinfo{journal}{Physics Letters A}
  \textbf{\bibinfo{volume}{374}}, \bibinfo{pages}{4784} (\bibinfo{year}{2010}).

\bibitem[{\citenamefont{Berman and Kezerashvili}(2012)}]{c:Berman2012}
\bibinfo{author}{\bibfnamefont{O.~L.} \bibnamefont{Berman}} \bibnamefont{and}
  \bibinfo{author}{\bibfnamefont{R.~Y.} \bibnamefont{Kezerashvili}},
  \bibinfo{journal}{Journal of Physics: Condensed Matter}
  \textbf{\bibinfo{volume}{24}}, \bibinfo{pages}{015305}
  (\bibinfo{year}{2012}).

\bibitem[{\citenamefont{Arefinia and Asgari}(2013)}]{c:Arefinia2013}
\bibinfo{author}{\bibfnamefont{Z.}~\bibnamefont{Arefinia}} \bibnamefont{and}
  \bibinfo{author}{\bibfnamefont{A.}~\bibnamefont{Asgari}},
  \bibinfo{journal}{Physica E: Low-Dimensional Systems and Nanostructures}
  \textbf{\bibinfo{volume}{54}}, \bibinfo{pages}{34} (\bibinfo{year}{2013}),
  ISSN \bibinfo{issn}{13869477}.

\bibitem[{\citenamefont{Qin et~al.}(2014)\citenamefont{Qin, Wang, Huang, Long,
  Wang, and Lu}}]{c:Qin2014}
\bibinfo{author}{\bibfnamefont{C.}~\bibnamefont{Qin}},
  \bibinfo{author}{\bibfnamefont{B.}~\bibnamefont{Wang}},
  \bibinfo{author}{\bibfnamefont{H.}~\bibnamefont{Huang}},
  \bibinfo{author}{\bibfnamefont{H.}~\bibnamefont{Long}},
  \bibinfo{author}{\bibfnamefont{K.}~\bibnamefont{Wang}}, \bibnamefont{and}
  \bibinfo{author}{\bibfnamefont{P.}~\bibnamefont{Lu}},
  \bibinfo{journal}{Optics Express} \textbf{\bibinfo{volume}{22}},
  \bibinfo{pages}{25324} (\bibinfo{year}{2014}).

\bibitem[{\citenamefont{Chaves et~al.}(2015)\citenamefont{Chaves, Peres, and
  Pinheiro}}]{PhysRevB.92.195425}
\bibinfo{author}{\bibfnamefont{A.~J.} \bibnamefont{Chaves}},
  \bibinfo{author}{\bibfnamefont{N.~M.~R.} \bibnamefont{Peres}},
  \bibnamefont{and} \bibinfo{author}{\bibfnamefont{F.~A.}
  \bibnamefont{Pinheiro}}, \bibinfo{journal}{Phys. Rev. B}
  \textbf{\bibinfo{volume}{92}}, \bibinfo{pages}{195425}
  (\bibinfo{year}{2015}).

\bibitem[{\citenamefont{Hajian et~al.}(2013)\citenamefont{Hajian, Soltani-Vala,
  and Kalafi}}]{c:Hajian2013}
\bibinfo{author}{\bibfnamefont{H.}~\bibnamefont{Hajian}},
  \bibinfo{author}{\bibfnamefont{A.}~\bibnamefont{Soltani-Vala}},
  \bibnamefont{and} \bibinfo{author}{\bibfnamefont{M.}~\bibnamefont{Kalafi}},
  \bibinfo{journal}{Optics Communications} \textbf{\bibinfo{volume}{292}},
  \bibinfo{pages}{149 } (\bibinfo{year}{2013}).

\bibitem[{\citenamefont{El-Naggar}(2015)}]{c:El-Naggar2015}
\bibinfo{author}{\bibfnamefont{S.~A.} \bibnamefont{El-Naggar}},
  \bibinfo{journal}{Optical and Quantum Electronics}
  \textbf{\bibinfo{volume}{47}}, \bibinfo{pages}{1627} (\bibinfo{year}{2015}).

\bibitem[{\citenamefont{Al-sheqefi and Belhadj}(2015)}]{c:Al-sheqefi2015}
\bibinfo{author}{\bibfnamefont{F.}~\bibnamefont{Al-sheqefi}} \bibnamefont{and}
  \bibinfo{author}{\bibfnamefont{W.}~\bibnamefont{Belhadj}},
  \bibinfo{journal}{Superlattices and Microstructures}
  \textbf{\bibinfo{volume}{88}}, \bibinfo{pages}{127} (\bibinfo{year}{2015}).

\bibitem[{\citenamefont{Madani and Entezar}(2013)}]{c:Madani2013}
\bibinfo{author}{\bibfnamefont{A.}~\bibnamefont{Madani}} \bibnamefont{and}
  \bibinfo{author}{\bibfnamefont{S.~R.} \bibnamefont{Entezar}},
  \bibinfo{journal}{Physica B: Condensed Matter}
  \textbf{\bibinfo{volume}{431}}, \bibinfo{pages}{1 } (\bibinfo{year}{2013}).

\bibitem[{\citenamefont{Cheng et~al.}(2015{\natexlab{a}})\citenamefont{Cheng,
  Chen, Yu, and Hsueh}}]{c:Cheng2015}
\bibinfo{author}{\bibfnamefont{Y.~H.} \bibnamefont{Cheng}},
  \bibinfo{author}{\bibfnamefont{C.}~\bibnamefont{Chen}},
  \bibinfo{author}{\bibfnamefont{K.~Y.} \bibnamefont{Yu}}, \bibnamefont{and}
  \bibinfo{author}{\bibfnamefont{W.~J.} \bibnamefont{Hsueh}},
  \bibinfo{journal}{Optics Express} \textbf{\bibinfo{volume}{23}},
  \bibinfo{pages}{28755} (\bibinfo{year}{2015}{\natexlab{a}}).

\bibitem[{\citenamefont{Kaipa et~al.}(2012)\citenamefont{Kaipa, Yakovlev,
  Hanson, Padooru, Medina, and Mesa}}]{c:Kaipa2012}
\bibinfo{author}{\bibfnamefont{C.~S.~R.} \bibnamefont{Kaipa}},
  \bibinfo{author}{\bibfnamefont{A.~B.} \bibnamefont{Yakovlev}},
  \bibinfo{author}{\bibfnamefont{G.~W.} \bibnamefont{Hanson}},
  \bibinfo{author}{\bibfnamefont{Y.~R.} \bibnamefont{Padooru}},
  \bibinfo{author}{\bibfnamefont{F.}~\bibnamefont{Medina}}, \bibnamefont{and}
  \bibinfo{author}{\bibfnamefont{F.}~\bibnamefont{Mesa}},
  \bibinfo{journal}{Physical Review B - Condensed Matter and Materials Physics}
  \textbf{\bibinfo{volume}{85}}, \bibinfo{pages}{4} (\bibinfo{year}{2012}).

\bibitem[{\citenamefont{Xu et~al.}(2013)\citenamefont{Xu, Chen, Wu, Wang, Teng,
  Zhang, and Bao}}]{c:Xu-experim}
\bibinfo{author}{\bibfnamefont{Z.}~\bibnamefont{Xu}},
  \bibinfo{author}{\bibfnamefont{C.}~\bibnamefont{Chen}},
  \bibinfo{author}{\bibfnamefont{S.~Q.~Y.} \bibnamefont{Wu}},
  \bibinfo{author}{\bibfnamefont{B.}~\bibnamefont{Wang}},
  \bibinfo{author}{\bibfnamefont{J.}~\bibnamefont{Teng}},
  \bibinfo{author}{\bibfnamefont{C.}~\bibnamefont{Zhang}}, \bibnamefont{and}
  \bibinfo{author}{\bibfnamefont{Q.}~\bibnamefont{Bao}},
  \emph{\bibinfo{title}{Graphene-polymer multilayer heterostructure for
  terahertz metamaterials}} (\bibinfo{year}{2013}).

\bibitem[{\citenamefont{Sreekanth et~al.}(2012)\citenamefont{Sreekanth, Zeng,
  Shang, Yong, and Yu}}]{c:Sreekanth2012}
\bibinfo{author}{\bibfnamefont{K.~V.} \bibnamefont{Sreekanth}},
  \bibinfo{author}{\bibfnamefont{S.}~\bibnamefont{Zeng}},
  \bibinfo{author}{\bibfnamefont{J.}~\bibnamefont{Shang}},
  \bibinfo{author}{\bibfnamefont{K.-T.} \bibnamefont{Yong}}, \bibnamefont{and}
  \bibinfo{author}{\bibfnamefont{T.}~\bibnamefont{Yu}},
  \bibinfo{journal}{Scientific reports} \textbf{\bibinfo{volume}{2}},
  \bibinfo{pages}{737} (\bibinfo{year}{2012}).

\bibitem[{\citenamefont{Fan et~al.}(2013)\citenamefont{Fan, Wei, Li, Chen, and
  Soukoulis}}]{c:Fan_plasmon2013}
\bibinfo{author}{\bibfnamefont{Y.}~\bibnamefont{Fan}},
  \bibinfo{author}{\bibfnamefont{Z.}~\bibnamefont{Wei}},
  \bibinfo{author}{\bibfnamefont{H.}~\bibnamefont{Li}},
  \bibinfo{author}{\bibfnamefont{H.}~\bibnamefont{Chen}}, \bibnamefont{and}
  \bibinfo{author}{\bibfnamefont{C.~M.} \bibnamefont{Soukoulis}},
  \bibinfo{journal}{Physical Review B} \textbf{\bibinfo{volume}{88}},
  \bibinfo{pages}{1} (\bibinfo{year}{2013}), \eprint{1311.7037}.

\bibitem[{\citenamefont{Wang et~al.}(2012)\citenamefont{Wang, Zhang,
  Garc{\'{i}}a-Vidal, Yuan, and Teng}}]{c:Wang_plasmon2012}
\bibinfo{author}{\bibfnamefont{B.}~\bibnamefont{Wang}},
  \bibinfo{author}{\bibfnamefont{X.}~\bibnamefont{Zhang}},
  \bibinfo{author}{\bibfnamefont{F.~J.} \bibnamefont{Garc{\'{i}}a-Vidal}},
  \bibinfo{author}{\bibfnamefont{X.}~\bibnamefont{Yuan}}, \bibnamefont{and}
  \bibinfo{author}{\bibfnamefont{J.}~\bibnamefont{Teng}},
  \bibinfo{journal}{Physical Review Letters} \textbf{\bibinfo{volume}{109}},
  \bibinfo{pages}{1} (\bibinfo{year}{2012}).

\bibitem[{\citenamefont{Fan et~al.}(2014)\citenamefont{Fan, Wang, Huang, Wang,
  Long, and Lu}}]{c:Fan_Bloch2014}
\bibinfo{author}{\bibfnamefont{Y.}~\bibnamefont{Fan}},
  \bibinfo{author}{\bibfnamefont{B.}~\bibnamefont{Wang}},
  \bibinfo{author}{\bibfnamefont{H.}~\bibnamefont{Huang}},
  \bibinfo{author}{\bibfnamefont{K.}~\bibnamefont{Wang}},
  \bibinfo{author}{\bibfnamefont{H.}~\bibnamefont{Long}}, \bibnamefont{and}
  \bibinfo{author}{\bibfnamefont{P.}~\bibnamefont{Lu}},
  \bibinfo{journal}{Optics Letters} \textbf{\bibinfo{volume}{39}},
  \bibinfo{pages}{6827} (\bibinfo{year}{2014}).

\bibitem[{\citenamefont{Wang et~al.}(2015{\natexlab{a}})\citenamefont{Wang,
  Qin, Wang, Ke, Long, Wang, and Lu}}]{c:Wang_Rabi2015}
\bibinfo{author}{\bibfnamefont{F.}~\bibnamefont{Wang}},
  \bibinfo{author}{\bibfnamefont{C.}~\bibnamefont{Qin}},
  \bibinfo{author}{\bibfnamefont{B.}~\bibnamefont{Wang}},
  \bibinfo{author}{\bibfnamefont{S.}~\bibnamefont{Ke}},
  \bibinfo{author}{\bibfnamefont{H.}~\bibnamefont{Long}},
  \bibinfo{author}{\bibfnamefont{K.}~\bibnamefont{Wang}}, \bibnamefont{and}
  \bibinfo{author}{\bibfnamefont{P.}~\bibnamefont{Lu}},
  \bibinfo{journal}{Optics Express} \textbf{\bibinfo{volume}{23}},
  \bibinfo{pages}{31136} (\bibinfo{year}{2015}{\natexlab{a}}).

\bibitem[{\citenamefont{Bludov et~al.}(2015)\citenamefont{Bludov, Smirnova,
  Kivshar, Peres, and Vasilevskiy}}]{c:Bludov_solitons2015}
\bibinfo{author}{\bibfnamefont{Y.~V.} \bibnamefont{Bludov}},
  \bibinfo{author}{\bibfnamefont{D.~A.} \bibnamefont{Smirnova}},
  \bibinfo{author}{\bibfnamefont{Y.~S.} \bibnamefont{Kivshar}},
  \bibinfo{author}{\bibfnamefont{N.~M.~R.} \bibnamefont{Peres}},
  \bibnamefont{and} \bibinfo{author}{\bibfnamefont{M.~I.}
  \bibnamefont{Vasilevskiy}}, \bibinfo{journal}{Physical Review B}
  \textbf{\bibinfo{volume}{91}}, \bibinfo{pages}{045424}
  (\bibinfo{year}{2015}).

\bibitem[{\citenamefont{Wang et~al.}(2015{\natexlab{b}})\citenamefont{Wang,
  Wang, Long, Wang, and Lu}}]{c:Wang_solitons2015}
\bibinfo{author}{\bibfnamefont{Z.}~\bibnamefont{Wang}},
  \bibinfo{author}{\bibfnamefont{B.}~\bibnamefont{Wang}},
  \bibinfo{author}{\bibfnamefont{H.}~\bibnamefont{Long}},
  \bibinfo{author}{\bibfnamefont{K.}~\bibnamefont{Wang}}, \bibnamefont{and}
  \bibinfo{author}{\bibfnamefont{P.}~\bibnamefont{Lu}},
  \bibinfo{journal}{Optics Express} \textbf{\bibinfo{volume}{23}},
  \bibinfo{pages}{32679} (\bibinfo{year}{2015}{\natexlab{b}}).

\bibitem[{\citenamefont{Sreekanth et~al.}(2013)\citenamefont{Sreekanth, Zeng,
  Yong, and Yu}}]{c:Sreekanth_biosensor2013}
\bibinfo{author}{\bibfnamefont{K.~V.} \bibnamefont{Sreekanth}},
  \bibinfo{author}{\bibfnamefont{S.}~\bibnamefont{Zeng}},
  \bibinfo{author}{\bibfnamefont{K.~T.} \bibnamefont{Yong}}, \bibnamefont{and}
  \bibinfo{author}{\bibfnamefont{T.}~\bibnamefont{Yu}},
  \bibinfo{journal}{Sensors and Actuators, B: Chemical}
  \textbf{\bibinfo{volume}{182}}, \bibinfo{pages}{424} (\bibinfo{year}{2013}).

\bibitem[{\citenamefont{Liu et~al.}(2014)\citenamefont{Liu, Liu, Wang, Wang,
  and Li}}]{c:Liu2014}
\bibinfo{author}{\bibfnamefont{J.-T.} \bibnamefont{Liu}},
  \bibinfo{author}{\bibfnamefont{N.-H.} \bibnamefont{Liu}},
  \bibinfo{author}{\bibfnamefont{H.}~\bibnamefont{Wang}},
  \bibinfo{author}{\bibfnamefont{T.-B.} \bibnamefont{Wang}}, \bibnamefont{and}
  \bibinfo{author}{\bibfnamefont{X.-J.} \bibnamefont{Li}},
  \bibinfo{journal}{Physica B: Condensed Matter}
  \textbf{\bibinfo{volume}{452}}, \bibinfo{pages}{66} (\bibinfo{year}{2014}).

\bibitem[{\citenamefont{Sensale-Rodriguez
  et~al.}(2012)\citenamefont{Sensale-Rodriguez, Yan, Kelly, Fang, Tahy, Hwang,
  Jena, Liu, and Xing}}]{c:Sensale-Rodriguez2012}
\bibinfo{author}{\bibfnamefont{B.}~\bibnamefont{Sensale-Rodriguez}},
  \bibinfo{author}{\bibfnamefont{R.}~\bibnamefont{Yan}},
  \bibinfo{author}{\bibfnamefont{M.~M.} \bibnamefont{Kelly}},
  \bibinfo{author}{\bibfnamefont{T.}~\bibnamefont{Fang}},
  \bibinfo{author}{\bibfnamefont{K.}~\bibnamefont{Tahy}},
  \bibinfo{author}{\bibfnamefont{W.~S.} \bibnamefont{Hwang}},
  \bibinfo{author}{\bibfnamefont{D.}~\bibnamefont{Jena}},
  \bibinfo{author}{\bibfnamefont{L.}~\bibnamefont{Liu}}, \bibnamefont{and}
  \bibinfo{author}{\bibfnamefont{H.~G.} \bibnamefont{Xing}},
  \bibinfo{journal}{Nature communications} \textbf{\bibinfo{volume}{3}},
  \bibinfo{pages}{780} (\bibinfo{year}{2012}).

\bibitem[{\citenamefont{Yan et~al.}(2012)\citenamefont{Yan, Li, Chandra,
  Tulevski, Wu, Freitag, Zhu, Avouris, and Xia}}]{c:Yan2012}
\bibinfo{author}{\bibfnamefont{H.}~\bibnamefont{Yan}},
  \bibinfo{author}{\bibfnamefont{X.}~\bibnamefont{Li}},
  \bibinfo{author}{\bibfnamefont{B.}~\bibnamefont{Chandra}},
  \bibinfo{author}{\bibfnamefont{G.}~\bibnamefont{Tulevski}},
  \bibinfo{author}{\bibfnamefont{Y.}~\bibnamefont{Wu}},
  \bibinfo{author}{\bibfnamefont{M.}~\bibnamefont{Freitag}},
  \bibinfo{author}{\bibfnamefont{W.}~\bibnamefont{Zhu}},
  \bibinfo{author}{\bibfnamefont{P.}~\bibnamefont{Avouris}}, \bibnamefont{and}
  \bibinfo{author}{\bibfnamefont{F.}~\bibnamefont{Xia}},
  \bibinfo{journal}{Nature Nanotechnology} \textbf{\bibinfo{volume}{7}},
  \bibinfo{pages}{330} (\bibinfo{year}{2012}).

\bibitem[{\citenamefont{Iorsh et~al.}(2013)\citenamefont{Iorsh, Mukhin,
  Shadrivov, Belov, and Kivshar}}]{c:Iorsh2013}
\bibinfo{author}{\bibfnamefont{I.~V.} \bibnamefont{Iorsh}},
  \bibinfo{author}{\bibfnamefont{I.~S.} \bibnamefont{Mukhin}},
  \bibinfo{author}{\bibfnamefont{I.~V.} \bibnamefont{Shadrivov}},
  \bibinfo{author}{\bibfnamefont{P.~A.} \bibnamefont{Belov}}, \bibnamefont{and}
  \bibinfo{author}{\bibfnamefont{Y.~S.} \bibnamefont{Kivshar}},
  \bibinfo{journal}{Physical Review B} \textbf{\bibinfo{volume}{87}},
  \bibinfo{pages}{075416} (\bibinfo{year}{2013}).

\bibitem[{\citenamefont{Zhukovsky et~al.}(2014)\citenamefont{Zhukovsky,
  Andryieuski, Sipe, and Lavrinenko}}]{c:Zhukovsky2014}
\bibinfo{author}{\bibfnamefont{S.~V.} \bibnamefont{Zhukovsky}},
  \bibinfo{author}{\bibfnamefont{A.}~\bibnamefont{Andryieuski}},
  \bibinfo{author}{\bibfnamefont{J.~E.} \bibnamefont{Sipe}}, \bibnamefont{and}
  \bibinfo{author}{\bibfnamefont{A.~V.} \bibnamefont{Lavrinenko}},
  \bibinfo{journal}{Physical Review B} \textbf{\bibinfo{volume}{90}},
  \bibinfo{pages}{155429} (\bibinfo{year}{2014}).

\bibitem[{\citenamefont{Cheng et~al.}(2015{\natexlab{b}})\citenamefont{Cheng,
  Fu, Weng, Chen, Zeng, Feng, and Chen}}]{c:Cheng_metamaterial2015}
\bibinfo{author}{\bibfnamefont{M.}~\bibnamefont{Cheng}},
  \bibinfo{author}{\bibfnamefont{P.}~\bibnamefont{Fu}},
  \bibinfo{author}{\bibfnamefont{M.}~\bibnamefont{Weng}},
  \bibinfo{author}{\bibfnamefont{X.}~\bibnamefont{Chen}},
  \bibinfo{author}{\bibfnamefont{X.}~\bibnamefont{Zeng}},
  \bibinfo{author}{\bibfnamefont{S.}~\bibnamefont{Feng}}, \bibnamefont{and}
  \bibinfo{author}{\bibfnamefont{R.}~\bibnamefont{Chen}},
  \bibinfo{journal}{Journal of Physics D: Applied Physics}
  \textbf{\bibinfo{volume}{48}}, \bibinfo{pages}{285105}
  (\bibinfo{year}{2015}{\natexlab{b}}).

\bibitem[{\citenamefont{Vakil and Engheta}(2011)}]{c:Vakil2011}
\bibinfo{author}{\bibfnamefont{A.}~\bibnamefont{Vakil}} \bibnamefont{and}
  \bibinfo{author}{\bibfnamefont{N.}~\bibnamefont{Engheta}},
  \bibinfo{journal}{Science (New York, N.Y.)} \textbf{\bibinfo{volume}{332}},
  \bibinfo{pages}{1291} (\bibinfo{year}{2011}).

\bibitem[{\citenamefont{Rejaei and Khavasi}(2015)}]{KhavasiScatt}
\bibinfo{author}{\bibfnamefont{B.}~\bibnamefont{Rejaei}} \bibnamefont{and}
  \bibinfo{author}{\bibfnamefont{A.}~\bibnamefont{Khavasi}},
  \bibinfo{journal}{J. of Opt.} \textbf{\bibinfo{volume}{17}},
  \bibinfo{pages}{075002} (\bibinfo{year}{2015}).

\bibitem[{\citenamefont{Bludov et~al.}(2013{\natexlab{b}})\citenamefont{Bludov,
  Ferreira, Peres, and Vasilevskiy}}]{c:primer}
\bibinfo{author}{\bibfnamefont{Y.~V.} \bibnamefont{Bludov}},
  \bibinfo{author}{\bibfnamefont{A.}~\bibnamefont{Ferreira}},
  \bibinfo{author}{\bibfnamefont{N.~M.~R.} \bibnamefont{Peres}},
  \bibnamefont{and} \bibinfo{author}{\bibfnamefont{M.~I.}
  \bibnamefont{Vasilevskiy}}, \bibinfo{journal}{International Journal of Modern
  Physics B} \textbf{\bibinfo{volume}{27}}, \bibinfo{pages}{1341001}
  (\bibinfo{year}{2013}{\natexlab{b}}).

\bibitem[{Not()}]{Note1}
\bibinfo{note}{Here the parity is considered with respect to the middle of the
  dielectric slab between the graphene sheets, i.e. the plane
  $z=\left(m+1/2\right)d$}.

\bibitem[{\citenamefont{Chang et~al.}(1974)\citenamefont{Chang, Esaki, and
  Tsu}}]{Esaki}
\bibinfo{author}{\bibfnamefont{L.~L.} \bibnamefont{Chang}},
  \bibinfo{author}{\bibfnamefont{L.}~\bibnamefont{Esaki}}, \bibnamefont{and}
  \bibinfo{author}{\bibfnamefont{R.}~\bibnamefont{Tsu}},
  \bibinfo{journal}{Appl. Phys. Lett.} \textbf{\bibinfo{volume}{24}},
  \bibinfo{pages}{593} (\bibinfo{year}{1974}).

\end{thebibliography}

\end{document}